\documentclass[11pt,onecolumn]{article}
\usepackage[latin1]{inputenc}
\usepackage{epsfig}
\usepackage[english]{babel}
\usepackage{graphicx}
\usepackage{subfigure}
\usepackage{textcomp}
\usepackage{amsmath}
\usepackage{amsfonts}
\usepackage{amssymb}
\usepackage{float}
\usepackage{xcolor}
\usepackage{abstract}
\usepackage{array,multirow}
\usepackage{authblk}
\usepackage{cite}

\title{\bfseries Characterization of ionization injection in gas mixtures irradiated by sub-petawatt class laser pulses }

\author{$\rm A. ~Zhidkov^{1,2,3}$}
\author{$\rm N. ~Pathak^{1,2}$\thanks{naveenpathak@sanken.osaka-u.ac.jp}}
\author{$\rm J. K.~Koga^{3}$}
\author{$\rm K. Huang^{2,3}$}
\author{$\rm M. Kando^{2,3}$}
\author{$\rm T. ~Hosokai^{1,2}$}

\affil{$\rm ^{1}$ \small {\em Institute of Scientific and Industrial Research (ISIR), Osaka University, Mihogaoka 8-1, Ibaraki, Osaka, 567-0047, Japan.}}
\affil{$\rm ^{2}$\small {\em Laser Accelerator R$\&$D , Innovative Light Sources Division, RIKEN SPring-8 Center, 1-1-1, Kouto, Sayo-cho, Sayo-gun, Hyogo, 679-5148, Japan}}
\affil{$\rm ^{3}$\small {\em Kansai Photon Science Institute, National Institutes for Quantum and Radiological Science and Technology, Kizugawa, Kyoto 619-0215, Japan}}

\date{}
\begin{document} 
\maketitle

\begin{abstract}
Effects of ionization injection in low and high Z gas mixtures for the laser wake field acceleration of electrons are analyzed with the use of balance equations and particle-in-cell simulations via test probe particle trajectories in realistic plasma fields and direct simulations of charge loading during the ionization process. It is shown that electrons appearing at the maximum of laser pulse field after optical ionization are trapped in the first bucket of the laser pulse wake. Electrons, which are produced by optical field ionization at the front of laser pulse, propagate backwards; some of them are trapped in the second bucket, third bucket and so on. The efficiency of ionization injection is not high, several $pC/mm/bucket$. This injection becomes competitive with wave breaking injection at lower plasma density and over a rather narrow range of laser pulse intensity.
\end{abstract}

\newpage

\section*{Introduction}
Interest in laser wake field acceleration (LWFA) of electrons \cite{Tajima} has rapidly grown over recent decades owing to notable results achieved by several groups \cite{Mangels,Geddes,Faure,WPL,WPLeeman,Gonsalves}. Production of 8 GeV electron bunches, in few tens of centimeters long plasma channel, with the charge around several $pC$ \cite{Gonsalves} demonstrates that LWFA may become a valuable branch in the electron accelerator family. Electron self-injection in the acceleration part of the laser wake field is a fundamental process for single-stage electron laser wake field acceleration \cite{Buck,Esarey,Ohkubo} as well as for a plasma cathode in multi-stage LWFA \cite{ Golovin,Sakai}. Presently, electron self-injection is considered to be the only way to generate electron bunches with characteristics suitable for their further acceleration in laser pulse wakes \cite{Buck,Sakai}. Therefore, investigation of the various mechanisms of electron self-injection is particularly important in the development of staging in LWFA.

So far plasma wave breaking mechanisms have been considered as the primary sources of electron self-injection. Several mechanisms such as wave breaking at density ramps \cite{Buck,Ohkubo}, parametric resonances \cite{Oguchi}, frequency chirp \cite{VBandhu,NPathak}, relativistic wave breaking \cite{Sheng,Bulanov}, and wave breaking provoked by external sources \cite{Faure2} have been proposed and studied both theoretically and experimentally. Recently, another mechanism of electron self-injection has been proposed; it is ionization injection \cite{Pak,Clayton,Ossa,Couperus,Kamperidis,Zeng2,Zheng2,Mirzaie,Thaury,Tomassini,Zhao,Mchen,
Mchen2,Mchen3}, which essentially differs from the common wave breaking process. 

Ionization injection should occur in a low $Z$ gas with a high $Z$ dope; for example $He-N_{2}$, $He-Ne$ and so on. Optical field ionization of inner shell electrons of the high $Z$ dope in the vicinity of the maximum of the laser pulse field produces a number of low energy electrons moving with a phase different from that of the laser pulse wake \cite{Pak}. These electrons can be trapped and further accelerated. 

Separation of ionization injection from the wave breaking injection can be done in rather low density plasma, which requires high power laser pulses in order to reach the self-focusing regime for essential electron acceleration. A laser pulse should have power equal to 2-3 times the critical power for self-focusing, $P_{cr}=1.7\times 10^{-2}N_{cr}/N_{e} \ [TW]$ \cite{EEsarey2}, where $N_{e}$ is the plasma density and $N_{cr}$ is the critical density corresponding to the laser wavelength. This implies that for $N_{e} < 10^{18}cm^{-3}$ and $\lambda=0.8 \ \mu m$, laser pulses with power of the order of petawatt class is required.

The principle of ionization injection and acceleration in a running wake wave can be illustrated by the simplest field structure as follows. A constant electric field with a negative strength $-E$, lasting from $V_{g}t$ to $L+V_{g}t$, where $t$ is the time, $L$ is the length of the wave moving with velocity $V_{g}$. If the initial position of an electron in this field at $t=0$ is $x=0$ (rear side of the wave), the electron, in order to be further accelerated or to move with the wave, must have a velocity, $v_{e}$ , such that $v_{e}>V_{g}$. This is the typical condition for the wave breaking process. In contrast, if an electron at $t=0$ has position $x=L$ (front of the wave), it can be trapped even if its velocity is initially zero. For this to happen the time $t_{0}$, over which the electron velocity becomes equal to the group velocity of the wave, should be such that the electron position must still exceed the position of the wave rear, $x_{0}=V_{g}t_{0}$ :

\begin{equation}
\begin{aligned}
v_{g} &=\frac{p_{0}+eEt_{0}}{mc\sqrt{1+\left( \frac{p_{0}+eEt_{0}}{mc} \right)^{2}}} \\
v_{g}t_{0} &\leqslant L+\frac{mc^{2}}{eE}\left[\sqrt{1+\left(\frac{p_{0}+eEt_{0}}{mc} \right)^{2}} - 1 \right]
\end{aligned}
\end{equation}

where $p_{0}$ is the initial electron momentum. The solution of Eq.(1) gives a rather soft condition: $U\geqslant mc^{2}\left(1-\frac{1}{\gamma_{g}}\right)-p_{0}v_{g}$, where $U=eEL$ is the potential difference and $\gamma_{g}$ is the relativistic factor for $V_{g}$. 

In particular for a laser pulse wake, assuming $E=E_{L}\frac{\omega_{pl}}{\omega_{0}}$ \cite{EEsarey2}, $p_{0}=0$ and $\gamma_{g} \gg 1$, where $E_{L}$ is the strength of the laser pulse, $\omega_{pl}$ the plasma frequency, and $\omega_{0}$ the laser frequency, one can get the requirement for the trapping length $L > \frac{\lambda_{pl}}{2\pi a_{0}}$ with $\lambda_{pl}$  being the plasma wavelength and $a_{0}=\frac{eE_{L}}{mc\omega_{0}}$ \cite{EEsarey2}. The $t_{0}$ can be considered to be the injection time, $t_{0}=\frac{\gamma_{g}}{\omega_{pl}a_{0}}$. This time is not short; it is of the order of one picosecond even for this maximum plasma field. Moreover, an electron, appearing inside a laser pulse, is trapped by the laser pulse fields and is accelerated by the pulse for a rather long time acquiring momentum opposite to the laser propagation direction owing to the ponderomotive force. The soft condition for the length results in a stronger condition for ionization injection on the laser pulse length and plasma density: $c\tau < \lambda_{pl}-2L$, where $\tau$ is the pulse duration assuming that half of the laser pulse occupies half of the plasma wake wave. It is clear that ionization injection can occur for low plasma density and rather short laser pulses. This explains the similarity in the results of electron acceleration in pure $Ar$ and $He$ gases at high densities \cite{Kando}. However, for lower plasma density, several groups have reported a clear dependence of the results on the dope quantity \cite{Mirzaie}. Nevertheless, there is an alternative point of view that the dope may aid in the modulation of plasma density by low intensity laser pre-pulses \cite{Hosokai,Lemos} resulting in much better focus-ability of the laser pulses. All these facts require detailed analysis of the ionization injection and its efficiency via particle-in-cell simulation.

There are two main approaches in the calculation of plasma ionization in particle-in-cell methods. The first is the variable particle weight technique \cite{Zhidkov,Zhidkov2,Makito} in the Villasenor-Buneman method, and the second is the variable number of particles technique \cite{ Zhidkov3,Kato,Zhidkov4}. The first is easier practically and less time consuming while the second allows for satisfying of the initial conditions for the particles when they are born. Not obeying the initial conditions for newly produced particles may sometimes result in non-physical solutions, see for example Ref. \cite{Zhidkov3}. Unfortunately, ionization injection also requires obeying the initial conditions, $p_{L}=0$, where $p_{L}$ is the particle momentum in the direction of the laser pulse propagation, and the variable particle weight technique is not applicable in its simple form. However, the density of 'ionization' electrons is usually small compared to the plasma density, the parameter $\eta /Z \ll 1$ even for $\eta =1$ [here $\eta$ is the doping ratio and $Z$ is the dope element charge]. Direct use of the method of variable number of particles \cite{Kato} with the necessary accuracy becomes impractical also.  On the other hand the low density of 'ionization' electrons allows for the use of a perturbative approach in the solution of the problem of ionization injection.

According to the calculations in Ref. \cite{Makito}, propagation of a laser pulse in high $Z$ gas with optical field ionization runs similarly to that in a plasma without ionization, if the plasma density is properly determined. Again, the doping of a high $Z$ gas in a low $Z$ gas can be considered as a small perturbation for the plasma and laser field. It is clear that for such conditions ionization injection can be characterized by a few parameters: plasma density, plasma density surplus in the vicinity of the laser pulse maximum, laser pulse intensity, and, weakly, by the spatial distribution of the laser light. In the present paper analysis of ionization injection depending upon these parameters is performed in three parts. In the first part the kinetics of the ion states, including inner shells, is considered for $He-N_{2}$ and $He-Ne$ gas mixtures to understand the spatial distribution of the 'ionization' electron density. The second part is devoted to the investigation of 'ionization' electron trajectories in real wake fields depending on the initial position of the electrons. In the third part charge loading in the acceleration phase of laser pulse wake owing to the ionization injection for different doping concentrations and plasma density is investigated via a self-consistent particle-in-cell simulation with separated post-processing for 'wave breaking' and 'ionization' electrons. This allows for an estimation of the efficiency of the different mechanisms of electron self-injection in gas mixtures.

\section*{Charge states in $\rm He-N_{2}$ and $\rm He-Ne$ mixtures}
The ion charge distribution under the action of strong femtosecond laser fields can be calculated with the common balance equations. In the absence of recombination the set of equations is rather simple:  
\begin{equation}
\begin{aligned}
\frac{\partial N_{0}}{\partial t} &= -S_{0}N_{0} \\
\frac{\partial N_{z}}{\partial t} &= S_{z-1}N_{z-1}
\end{aligned}
\end{equation}

and,

\begin{equation}
\begin{aligned}
\frac{\partial N_{k}}{\partial t} &= S_{k-1}N_{k-1}-S_{k}N_{k} \\ 
k &= 1,....,Z-1 \nonumber
\end{aligned}
\end{equation}

where $N_{k}$ is the density of ions with charge $k$, $S_{k}$ is the optical field ionization rate, $Z$ is the nuclear charge of an element. There are several approximations for the ionization rate \cite{Popruzhenko}. We use the following form 

\begin{equation}
S_{k}=4 \omega_{A} g_{k}\left(I_{k}/Ry \right)^{5/2}\left(E_{A}/E_{L}\right) \exp\left(-\frac{2}{3}\left(I_{k}/Ry \right)^{3/2}\left(E_{A}/E_{L}\right) \right) \ [s^{-1}] \nonumber
\end{equation}

with ionization potentials $I_{k}$ for $He$, $N_{2}$, and $Ne$ ions taking data from the most comprehensive calculations and experiments \cite{NIST}. Here $E_{A}=m^{2}e^{5}/\hbar^{4}$, $\omega_{A}=me^{4}/\hbar^{3}$, $Ry=m^{2}e^{4}/2\hbar^{2}$, and $g_{k}$ is a factor of the order of unity. For a gas mixture Eq. (2) should be calculated for each gas species. In Eq. (2) we implicitly assume that the density of ions with charge $n$ for which $(I_{n}/Ry)^{3/2}E_{A}/E_{L} \ll 1$ is already negligibly small.

Numerical solution of Eq. (2) with the conservative condition $$\sum_{0}^{Z} N_{k}=constant$$ is not difficult. It is clear that there is a saturation level of ion charge owing to the exponential dependency of the ionization rate on the ionization potential. In $He$ gas such a saturation occurs at the front of a powerful laser pulse. In the presence of a high Z dope; ionization of the outer shells and inner shell of the ions is different, because the ionization potential of the inner shell is essentially higher than those for the outer shells by the parameter $\xi = [2Z/(Z-2)]^{2}$; for $Z \gg 1$ it becomes equal to 4. This means that after a plateau there may be a density jump in the vicinity of the maximum of the laser pulse field strength.

In Fig.1 and Fig.2 typical distributions of the electron density produced by a Gaussian laser pulse with different intensities and a duration of $30 fs$ (such a pulse duration is typical for most of the existing high lower laser facilities) are presented for $He-N_{2}$ and $He-Ne$ gas mixtures. To make the jump more visible the concentrations of the dopes is chosen to be $5\%$ for $N_{2}$ and $10\%$ for Ne. Figs.1 (a,b,c) show the evolution of the electron density along the laser axis. In the case of the nitrogen dope essentially a density jump in the vicinity of the laser pulse maximum (clear in insets) is seen already for $a_{0}=1.5$ and vanishes for $a_{0}=3.5$. Within a distance of about $1 \ \mu m$ near the pulse maximum the density growth $\bigtriangleup N$ is less than $1\%$ for all sets of $a_{0}$. With an increase of $a_{0}$ the density jump apparently shifts towards the front of the laser pulse. However, along the periphery of the laser pulse there is a density jump for $a_{0}=3.5$ as seen in Fig.1(d). Such a jump has a ring like shape. Further dynamics of such electrons have to be investigated as well. The higher ionization potential of Ne results in a higher $a_{0}$ necessary to have a density jump in the vicinity of maximum of the laser pulse as seen in Fig.2 (a,b,c) (see insets). In the case of $a_{0}=6$ the density growth reaches several percent and exceeds that for nitrogen. For $a_{0}>8$ the density jump vanishes in the vicinity of the laser pulse field maximum and shifts towards its front. This is important for further electron dynamics as described in the next session. Again, there is a ring shape density jump in the periphery of the laser pulse even for $a_{0}=10$ as shown in Fig. 2(d).

The distributions of 'ionization' electrons which can be extracted for Fig.1 and Fig.2 are very important in understanding the dynamics of ionization injection into the laser pulse wake. To clarify the physical picture of the dynamics of such electrons we perform an investigation of electron trajectories in the realistic laser wake fields.

\section*{Probing particles in particle-in-cell simulations}
To understand how the position of 'ionization' electron influences its further dynamics we perform tests with probe particles during the propagation of a laser pulse in plasma using the particle-in-cell method. The simulations are performed in 2D geometry, using the moving window technique. The window has size $(100 \times 200) \ \mu m^{2}$; the spatial grid resolution is $\lambda / 36$. Laser pulses with duration of $30 fs$ propagate in a uniform plasma with a density of $N_{e}=1\times 10^{18}cm^{-3}$. The laser pulse intensity is varied from $I=1\times 10^{19} Wcm^{-2}$ to $I=3\times 10^{19} Wcm^{-2}$ and the laser spot size is $w_{0}=10 \ \mu m$. Probe particles with small charge are placed near the maximum of the laser pulse after $140 \ \mu m$ of pulse propagation and start moving. There are five sets of the test probe particles, which are distributed around the peak of the laser pulse. Each set has six particles located at different positions in the transverse direction. Two sets of the test particles are placed in front side of the laser pulse, and two sets are placed at the rear side. One set of particles is placed at the peak of the pulse. These positions mimic well the appearance of 'ionization' particles with initially zero momenta.

The resulting trajectories of the probe electrons for laser pulse intensity $I=3\times 10^{19} Wcm^{-2}$ are presented in Fig. 3 (a-d) at $t=3.5 \ ps$. Particles situated at the maximum of the laser pulse have the trajectories shown in Fig.3(a). Initially particles are trapped by the laser pulse $v \times B$ force and move along the laser pulse propagation direction. Then the particles, having been overrun by the laser pulse, are taken up by the wake field and accelerated. After approximately one picosecond all these particles are accelerated above the phase velocity of the plasma wave and, therefore, are injected. The evolution of their momenta can be seen in Fig.4(a): after short motion in the laser pulse the probe particles are accelerated to over $100 \ MeV$ after several picoseconds. The relativistic factor for the phase velocity corresponding to the plasma density is $\gamma_{ph}=\sqrt{\frac{N_{cr}a_{0}}{N_{e}}} \approx 70$. The particles reach this energy after about $1.5 \ ps$. The effect of side scattering by the ponderomotive force is illustrated in Fig.4(b) by the transverse momenta of the particles. The effect exists but is small for the particles, the geometrical emittance for them is within $3 \times 10^{-3} \ rad$.

A very different picture emerges in the case where the initial positions of the particles are before the maximum of the laser pulse field as shown in Fig. 3(b). One can observe no trapping of these particles in the first bucket of the wake. However, one particle is trapped and accelerated in the second bucket. The fact of trapping for this particle can be proved by particle momentum shown in Fig.4(c) where the velocity of this particle exceeds the phase velocity of the wave. The emittance for this particle is not small being of the order of $0.06 \ rad$. Particles situated further from the pulse field maximum cannot be trapped even in the second bucket; a portion of them move to the third bucket and can be trapped there. This effect strongly depends on the longitudinal momentum acquired by the particle after the laser pulse has overrun it. The simulation shows that particles in the front of the laser pulse get a larger momentum directed counter to the laser pulse motion. Moreover, we observe no trapping in the first bucket of the probe particles situated at the maximum of the laser field in the case of lower pulse intensity $I=1 \times 10^{19} Wcm^{-2}$ with the other conditions being the same. For higher intensity, some particles situated behind the maximum of the pulse field get almost zero momentum and can be efficiently trapped, as shown in Fig.4(d). However, the number of such particles is not high as seen From Fig.1 and Fig.2 where the increase in the number of the 'ionization' electrons after the peak of the laser pulse is small or negligible compared to the whole plasma density. In contrast most of 'ionization' electrons are born before the maximum of the laser field. This may result not only in continuous injection of 'ionization' electrons in the first bucket but quite efficient injection of them into consequent buckets. Since the ionization injection is more efficient in the absence of wave breaking the number of buckets can be large, which results in a long length for the final electron bunches of the order of picoseconds.

\section*{Ionization injection in low density plasma}
Effects of ionization injection including charge loading can be calculated only with the use of the particle-in-cell method. It is apparent that ionization of plasma should be included. Upon considering a lower intensity laser pulse $a_{0}<10$ we neglect ion motion and consider ionization processes using this approximation. Again, there are two methods to calculate plasma ionization in the framework of particle-in-cell simulation. The first is by the variable particle weight (VPW) method. In this method, particles emulating electrons are initially empty and are filled with electrons in time according with Eq. (1) \cite{Makito}. This method is not time consuming. However, it has a disadvantage: it is impossible to apply arbitrary initial conditions for the electrons. For example, in the problem of light scattering from an ionization wave the fact that the variable particle weight method results in unphysical solutions is explained in detail in \cite{Zhidkov3}. Another method is the variable number of particles (VNP) \cite{Kato,Zhidkov4}. $N$ particles per a cell have the same weights $W$. The value $N \times W$ represents the total possible electron density in the plasma. Initially all particles are immobile. With time the number of movable particles increases in accordance with Eq. (1). A new particle is involved in the motion with the necessary initial conditions for its momentum. The disadvantage of this approach is that large number of particles are necessary to make a smoother spatial electron density distribution. In the case of ionization injection the density jump, $\bigtriangleup N$,  is rather small and the use of VNP requires a resolution better than $\bigtriangleup N/N_{e}$ or the number of particles $N> \frac{N_{e}}{\bigtriangleup N}$. 

There is an alternative approach allowing essential lessening of computing resources with a rather accurate evaluation of the ionization injection. The alternative is in the combination VNP and VPW.  For the case of $\bigtriangleup N/N_{e} \ll 1$ which is typical for experiments with doping it is reasonable to use two groups of electrons, see Fig.1(b). The first one serves as 'plasma' electrons. The dynamics of these electrons can be calculated both with the use of VPW or assuming a pre-ionized plasma \cite{Makito}. The second group consists of 'ionization' electrons for which the VNP method is applied. We use for this group $m$ slices with charges proportional to $\bigtriangleup N/m$ and with a motion parameter which switches from 'false' to 'true' when a $k$ immovable 'ionization' particle crosses (in the moving window) the vicinity of the laser field with strength $E_{Lk}$. With the motion parameter 'true' the 'ionization' particle is involved in self-consistent motion similar to 'plasma' particles. 

The simplest case is the single slice of 'ionization' particles which start moving when crossing the 'maximal' strength of the laser field. We perform such alternative particle-in-cell simulations in two dimensions for gas mixtures using $N_{e}$ and $\bigtriangleup N$ as parameters for a plasma irradiated by a $30 fs$ laser pulse with $\lambda=0.8 \ \mu m$ and intensity $I=3\times 10^{19} Wcm^{-2}$ focused to $w_{0}=10 \ \mu m$. The moving box has size $(100 \times 200) \ \mu m^{2}$ and the spatial grid resolution $\lambda/36$. The number of slices for 'ionization' was one and five. We use a uniform plasma with a fixed density for easier control of the parameters. According to Ref. \cite{Makito} such an approach does not essentially change the plasma field distributions even for pure high $Z$ gas. The transverse distribution of the charges of the 'ionization' particles is Gaussian with size equal $0.5w_{0}$ centered around the laser propagation axis.

Momentum distributions for 'plasma' and 'ionization' particles for a plasma with density $Ne=1\times 10^{18} cm^{-3}$ are shown in Fig. 5 (a-d) at $t=4.1 \ ps$ pulse propagation and for $\bigtriangleup N=10^{-2}N_{e}$. Such a density jump corresponds to $5\%$ of $N_{2}$ according to the ionization balance in the laser pulse field. According to 'plasma' particles, see Fig 5(a,b), the wave breaking injection is rather weak and occurs in the third and fourth buckets. As for ionization injection, Fig.5(c,d), it starts in the first bucket and gradually appears in the other buckets. The energy of the injected electrons in the first bucket reaches $150 \ MeV$ after $4 \ ps$ and increases. The electron energy in the second, third, and fourth bucket also increases similarly with a clear time delay. The evolution of the total energy distribution taking into account the particle weights is shown in Fig. 6(a-c). One can see two parts separated by a peak in the figures. The lower energy part originates from the wave breaking while the higher energy part is from pure ionization injection. Both parts are flat without characteristic features. However, the wave breaking injection provided a larger charge of accelerated electrons than the ionization injection as seen in Fig. 6(c). We also have to separate out different sorts of electrons from the ionization injection. There are many electrons involved in the initial acceleration by the pulse wake. These electrons acquire certain energies around up to tens of $MeV$. However, they are not finally trapped in any bucket and form a sort of clouds with very low emittance. The number of such electrons far exceeds the number of injected electrons. Another negative feature of ionization injection is the lengthening of the total electron bunch after electron injection in consequent buckets. Since the wave breaking process is not strong, as seen in Fig. 7 (a,b), the number of buckets is quite large. All these buckets will be filled with injected electrons which are further accelerated. Such beams may have durations of several picoseconds. 

The results presented in Fig. 5-7 have been obtained for a rather low concentration of high $Z$ gas. Fig. 8(a-d) illustrates what happens if the density of high $Z$ gas is increased by factor of 10. One can see that the ionization injection in this case becomes dominant. Moreover, it suppresses the electron self-injection by the wave breaking process as seen in Fig. 8(a,b). However, the maximal energy of the accelerated electrons is lower than that for the lower concentration of high $Z$ gas (also see Fig. 9). The most efficient injection and acceleration occur in the second bucket, which reflects upon the increase of the effect of the pre-accelerated but not injected electrons on the acceleration process. The number of such electrons essentially increases. It is important to note that the practical realization of such conditions requires much higher laser pulse power. Since now the electron density is formed by a high $Z$ gas, the power of the laser pulse should be equal to 2-3 times the critical power not for the maximum ionization plasma density but for the minimum, which is almost 10 times smaller. This is because as the laser pulse focuses the intensity is low enough so that initially the laser is only propagating in a singly ionized plasma. To insure relativistic focusing this requires higher laser power. For the present parameters $N_{2}$ gas requires a density of $N_{e}=1\times 10^{18} cm^{-3}$ for which the laser pulse power should be about petawatt levels. Otherwise the pulse diffraction will shorten the length of the higher density plasma. Working with high power may be critical to the ionization injection, however, this may restrict the process by itself.  For laser pulse intensities approaching $I=1 \times 10^{20} Wcm^{-2}$ relativistic wave breaking \cite{Sheng}, which occurs for $a_{0}\sqrt{2\left(\gamma_{ph}-1 \right)}$, along with transverse wave breaking \cite{Bulanov}, which occurs for $\lambda_{pl}a^{1/2}/w_{0} \gg 1$, may make the wave breaking injection dominant even for low density plasma.

The pure ionization injection case, which is at lower density than the previous cases, is illustrated in Fig. 10 (a-d). In a plasma with density $N_{e}=5\times 10^{17} cm^{-3}$ and $\bigtriangleup N=5 \times 10^{16} cm^{-3}$ irradiated by laser pulses with intensity $I=3 \times 10^{19} Wcm^{2}$ we observe no results of the wave breaking processes. It clearly seen in Fig.10 (a,b) where there is no visible self-injected plasma electrons. In contrast, one can see in Fig.10 (c,d) an efficient injection to the third bucket with continuous electron acceleration after $4 \ ps$ of laser pulse propagation. In the first and second buckets there are only pre-accelerated electrons. At this time, $t=4.1 \ ps$, it is impossible to say whether these electrons will be further accelerated or not. In Fig.11 spatial distributions of the total electron density and plasma electric field are shown. One can see a small channel-like structure appearing along the laser axis. We anticipate that this structure is formed by pre-accelerated electrons moving through the buckets. The evolution of the total distribution function is shown in Fig. 12. Even for $t=4.1 \ ps$ we observe no electrons with velocity exceeding the phase velocity of the plasma wave, which for this condition should be about $\gamma=100$. 

\section*{Conclusion}
In conclusion, we have characterized the effect of ionization injection for the laser wake field acceleration of electrons in gas mixtures using self-consistent particle-in-cell simulations. First, we applied the technique of test probe particles to investigate the condition of particle trapping in the acceleration phase of the wake field. Second, we performed two dimensional particle-in-cell simulations splitting 'plasma' electrons and ionization' electrons, which allowed us to investigate the entire process of electron pre-acceleration, trapping, and further acceleration.

Direct simulations have shown the efficiency of continuous ionization injection in a low density plasma with low and high $Z$ components. Electrons, generated by optical field ionization of the ion inner shell in the vicinity of the maximum of the laser pulse field, after their propagation inside the pulse can be pre-accelerated by the wake field to energies high enough for their further acceleration. These electrons can form a high energy, low emittance beam. The total charge of the beam depends on the matching between the laser intensity and the high $Z$ component. Electrons appearing before the maximum of laser pulse field due to ionization of inner shell cannot be trapped in the first bucket and finally form a cloud of pre-accelerated electrons or are injected in wake buckets far behind the laser pulse. The later process results in the formation of a long electron bunch with a duration of several picoseconds. Optimal matching can provide a density jump equal $\eta N_{e}/Z$ with $\eta$ being the concentration of the high $Z$ gas. With plasma density increase the ionization injection cannot compete with the wave breaking injection. Moreover, when the laser pulse length exceeds the plasma wavelength, the ionization injection vanishes. 

Typically in experiments the density of the high $Z$ gas is small being of the order of a few percent. For densities of ionization electrons $\bigtriangleup N=10^{15}-10^{16} cm^{-3}$ [$
(5-10) \ \%$ of high $Z$ dope] and a diameter of the injection section of about $5 \ \mu m$ (see Fig.1 and Fig.2) one can estimate the total charge per $mm$ of acceleration length to be $(40-400) \ pC/mm$. However, the low efficiency for trapped electrons, which is less than $10\%$ of the total pre-accelerated particles, essentially reduces the charge of accelerated electrons. The total charge may be considerably higher after ionization injection fills many wake buckets. However, such beams can be quite long with their duration being picoseconds. 

\section*{Acknowledgements}
This work is funded by the JST-MIRAI program grant no. JPMJMI17A1, and was partially supported by the ImPACT R$\&$D Program of Council for Science, Technology and Innovation (Cabinet Office, Government of Japan). We are grateful to Prof. Yuji Sano for encouragement and helpful discussions. We also acknowledge the use of Mini-K computing facility at SACLA, RIKEN, SPring-8 Center. 


\newpage
\begin{figure}[!tb]
\centering
\subfigure[]
{\includegraphics[width=7.0cm]{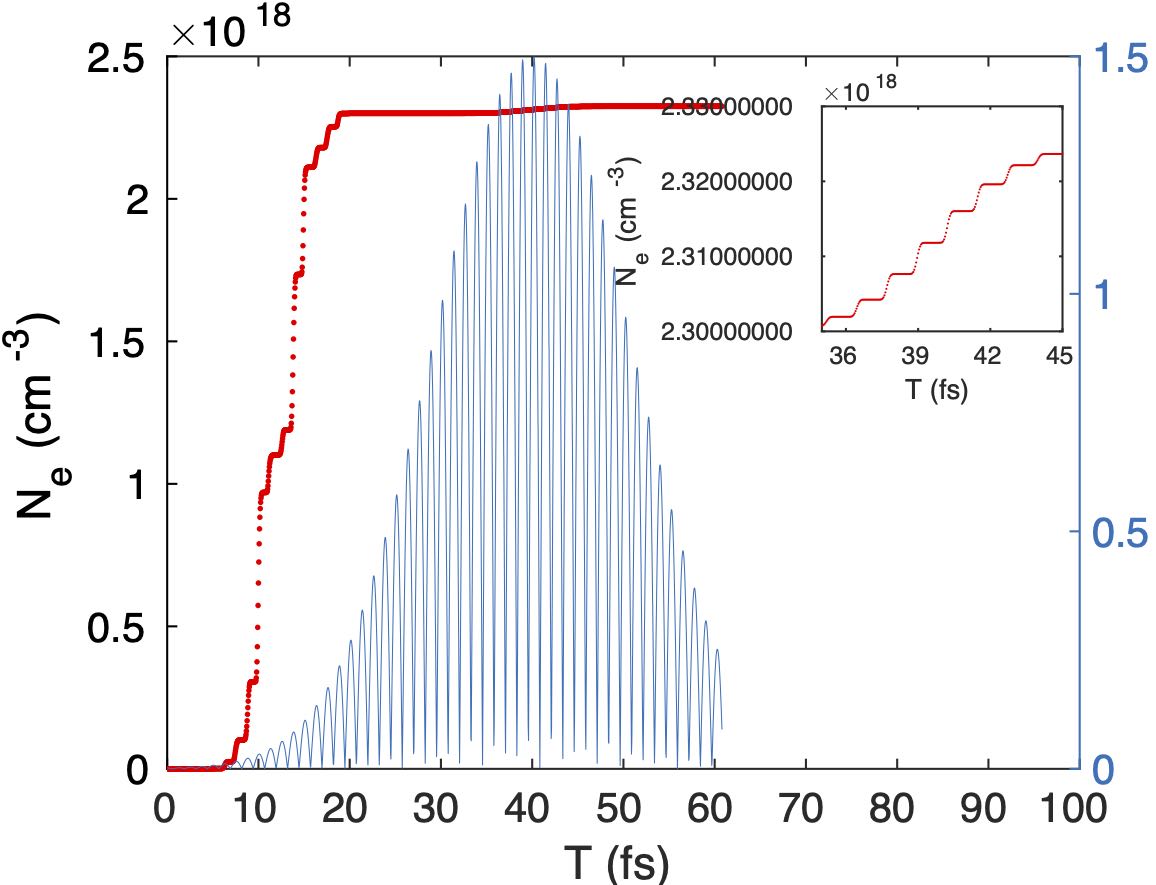}}
\hspace {0.3cm}
\subfigure[]
{\includegraphics[width=7.0cm]{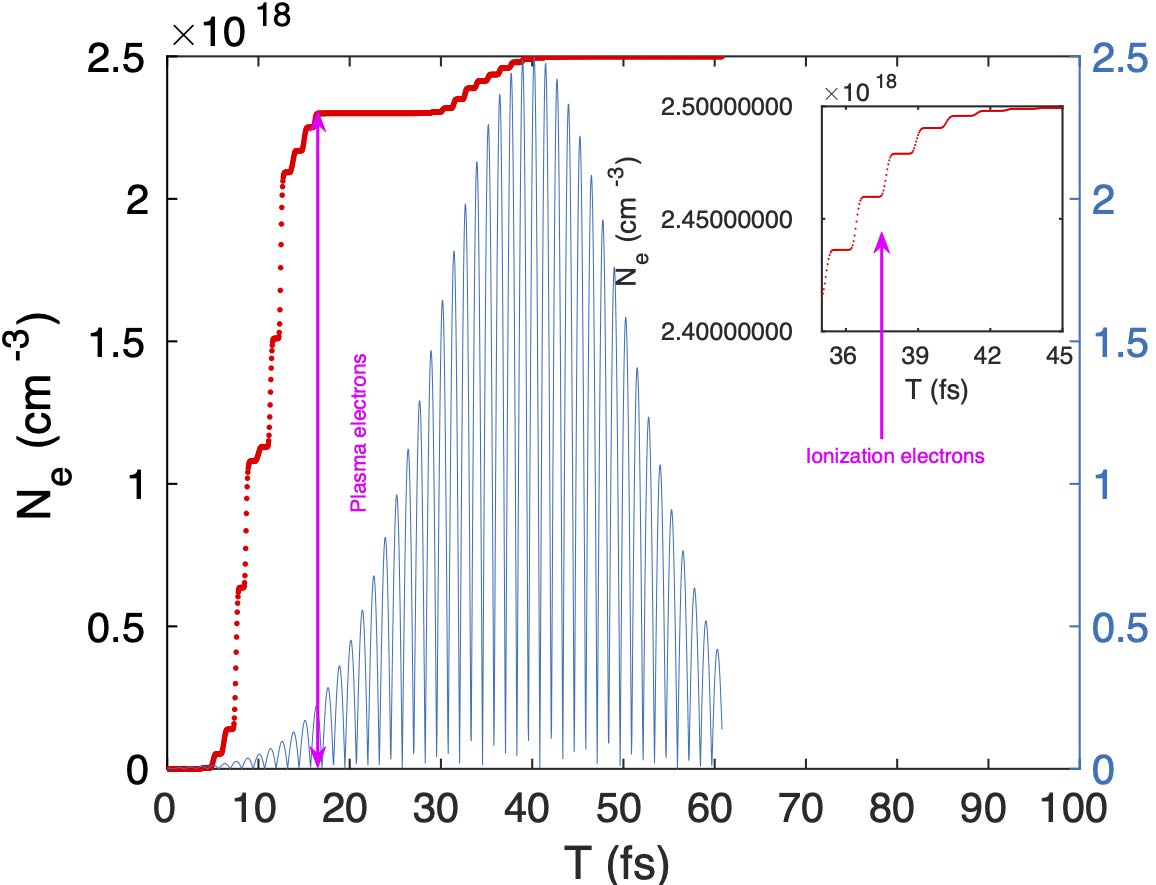}}
\subfigure[]
{\includegraphics[width=7.0cm]{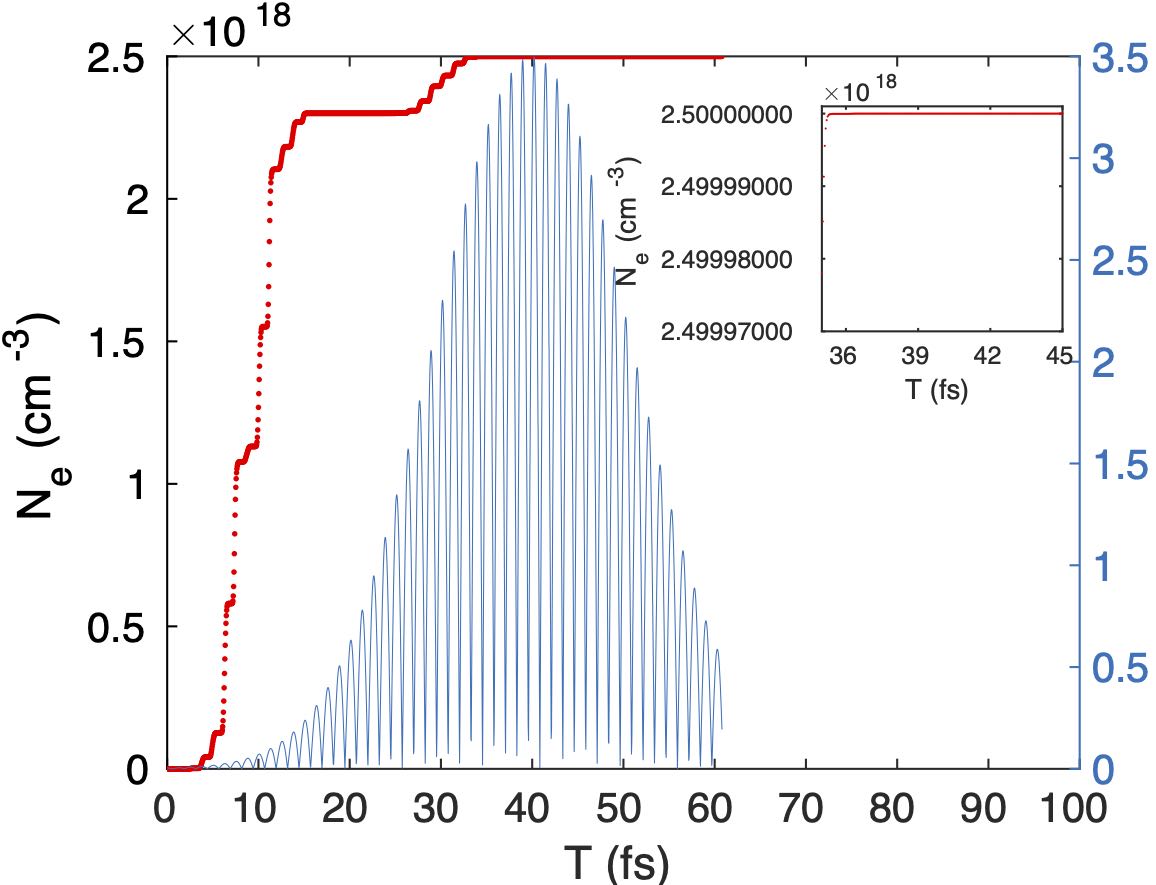}}
\hspace {0.3cm}
\subfigure[]
{\includegraphics[width=7.0cm]{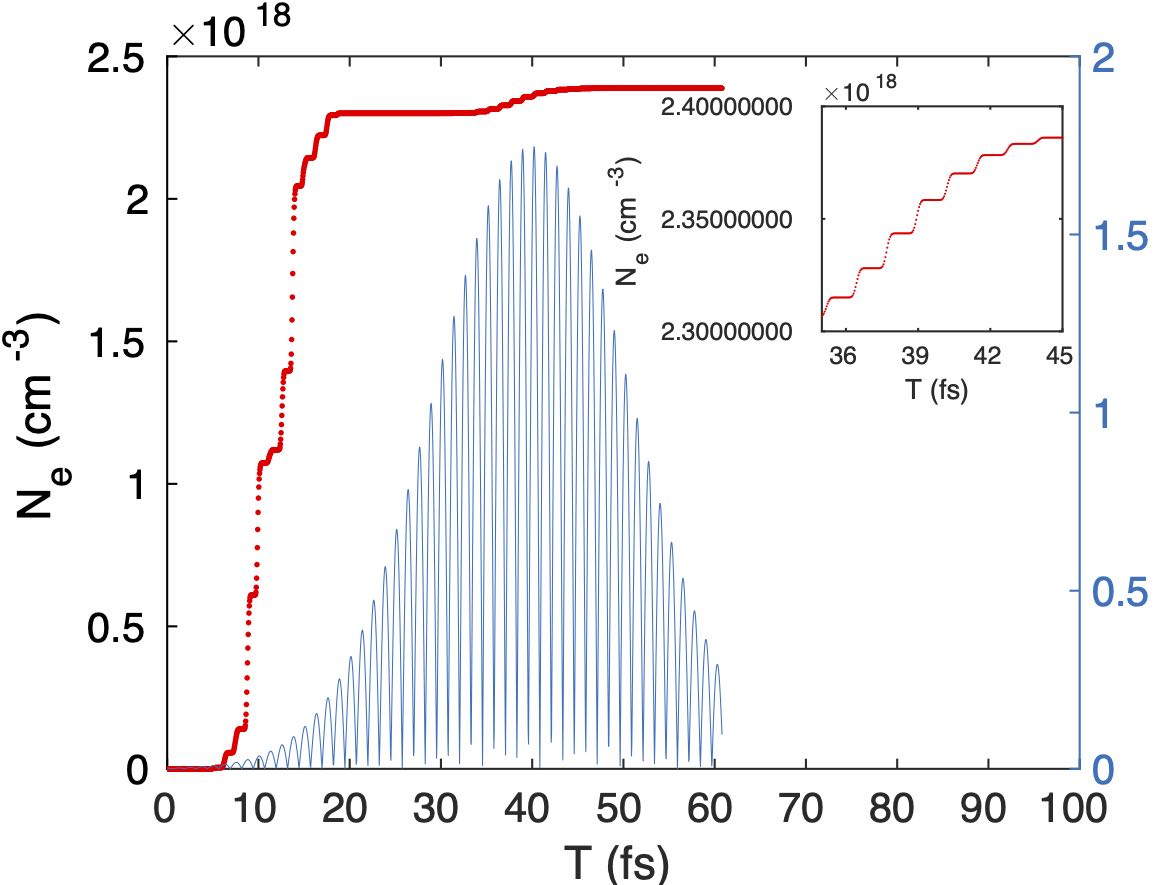}}
\caption{ Dynamics of the electron density in the gas mixture $He-N_{2}$ with $5\%$ of $N_{2}$ along the laser axis (a-c), and off-axis (d) depending on the laser pulse intensity: (a) $a_{0max}=1.5$, (b) $a_{0max}=2.5$, and (c,d) $a_{0max}=3.5$. Arrows in (b) shows the range of 'plasma' electrons and 'ionization' electrons. Insets show the electron densities around the peak of the laser pulse.}
\label{fig.1}
\end{figure}

\begin{figure}[!tb]
\centering
\subfigure[]
{\includegraphics[width=7.0cm]{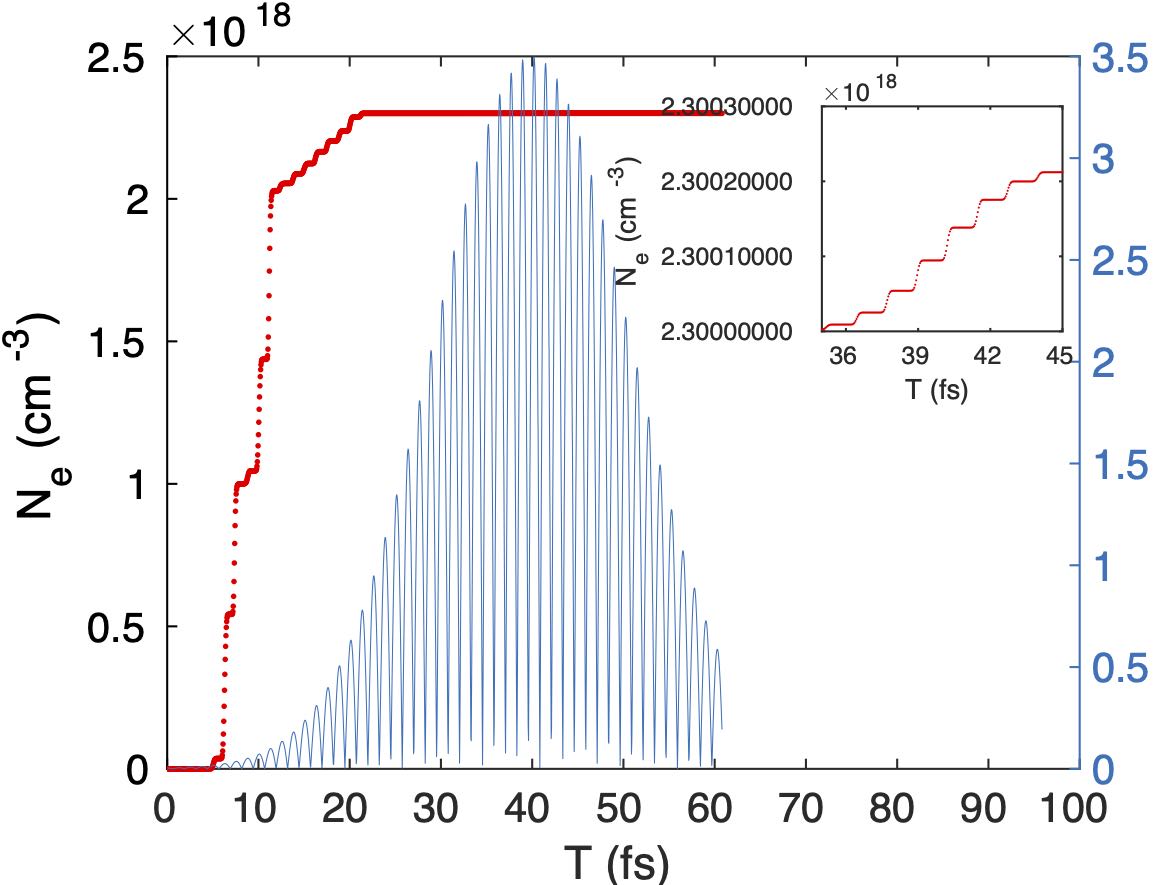}}
\hspace {0.3cm}
\subfigure[]
{\includegraphics[width=7.0cm]{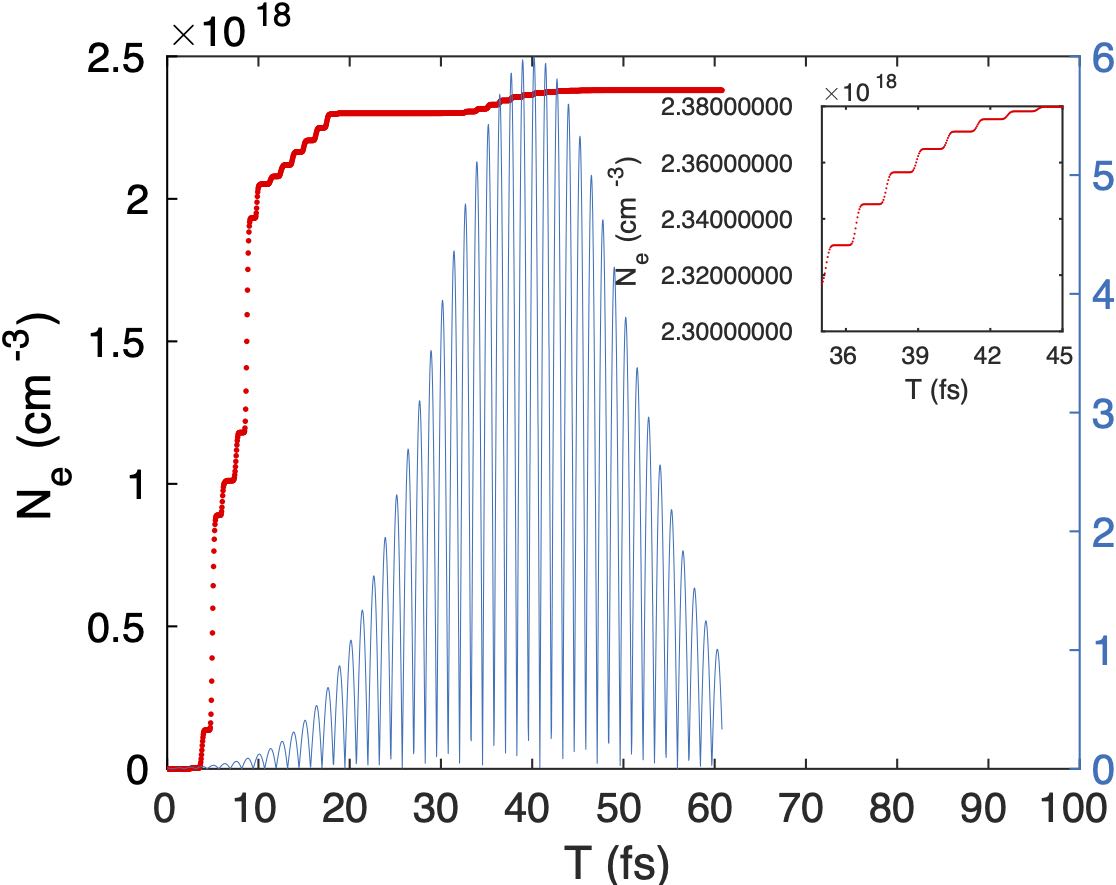}}
\subfigure[]
{\includegraphics[width=7.0cm]{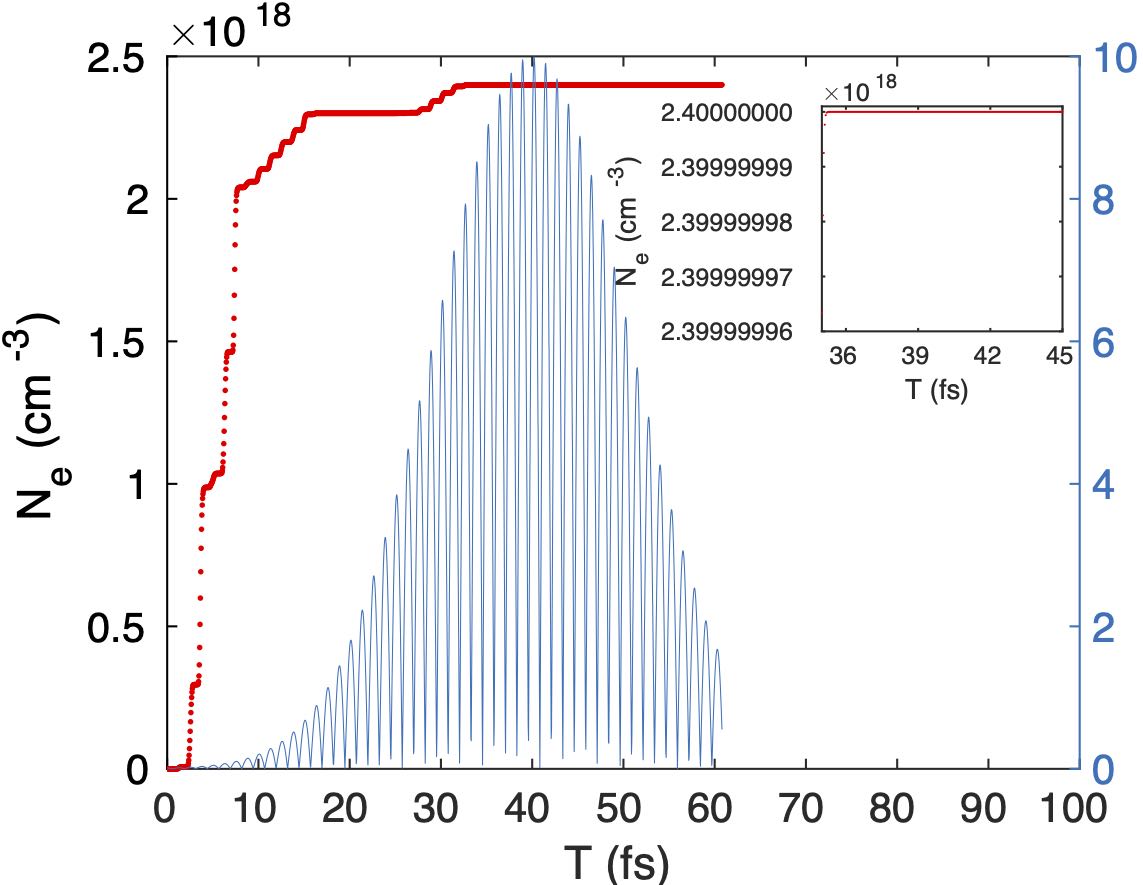}}
\hspace {0.3cm}
\subfigure[]
{\includegraphics[width=7.0cm]{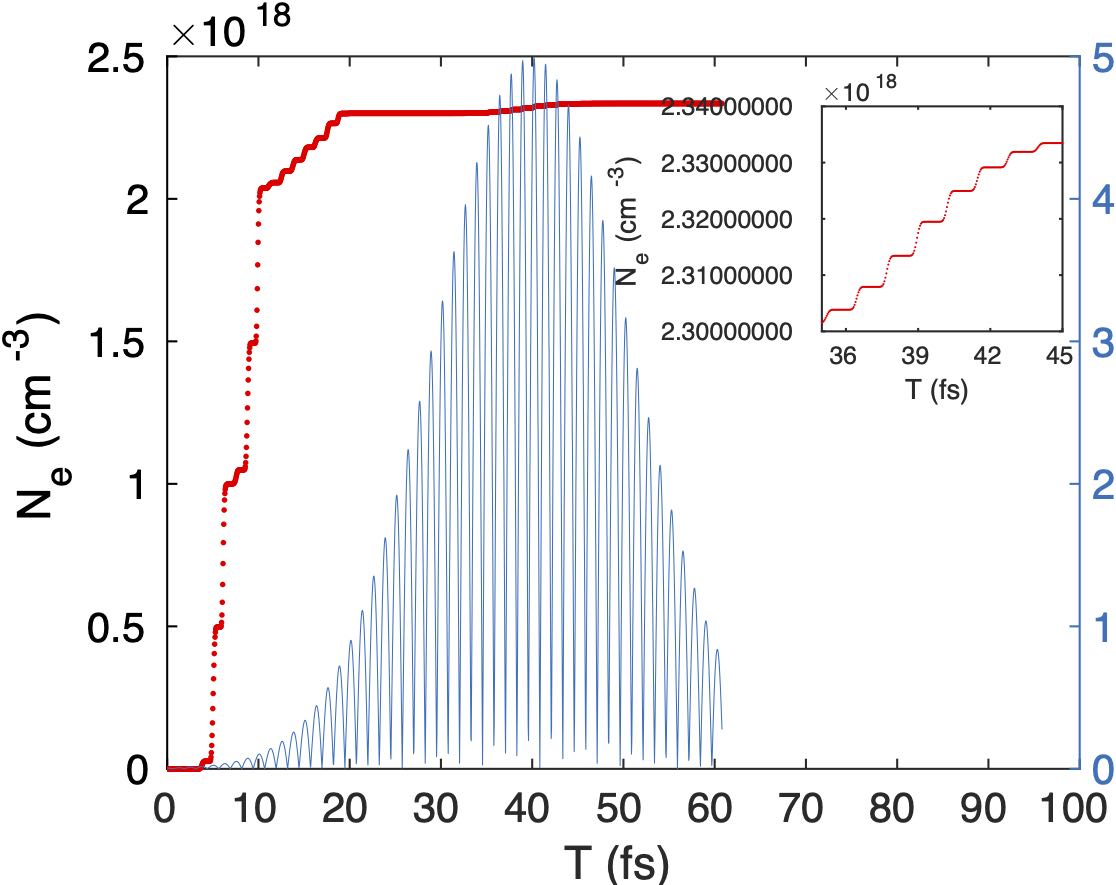}}
\caption{ Dynamics of the electron density in the gas mixture $He-Ne$ with $10\%$ of $Ne$ along the laser axis (a-c), and off-axis (d) depending on the laser pulse intensity: (a) $a_{0max}=3.5$, (b) $a_{0max}=6$, and (c,d) $a_{0max}=10$. Insets show the electron densities around the peak of the laser pulse.}
\label{fig.2}
\end{figure}

\begin{figure}[!tb]
\centering
\subfigure[]
{\includegraphics[width=7.0cm]{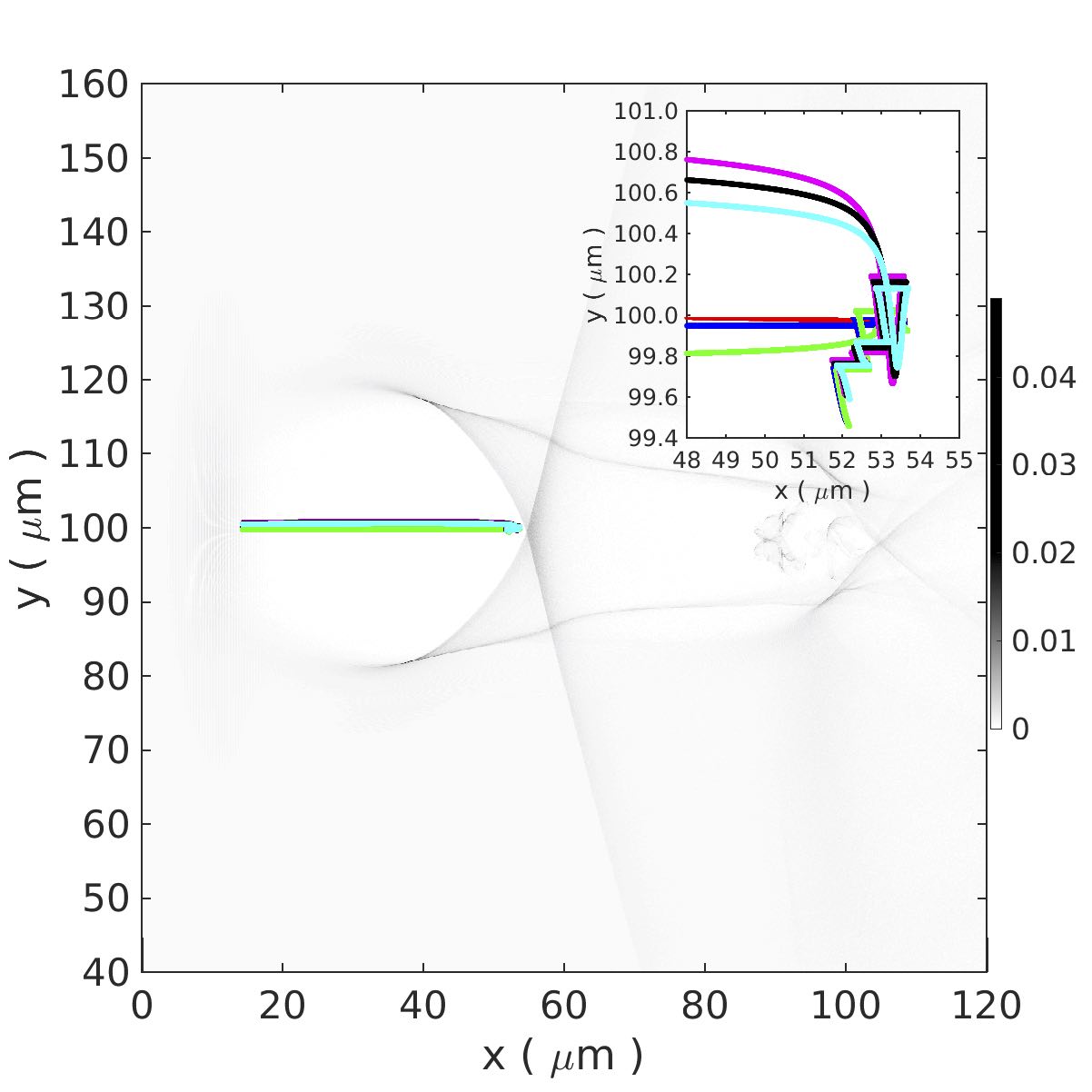}}
\hspace {0.3cm}
\subfigure[]
{\includegraphics[width=7.0cm]{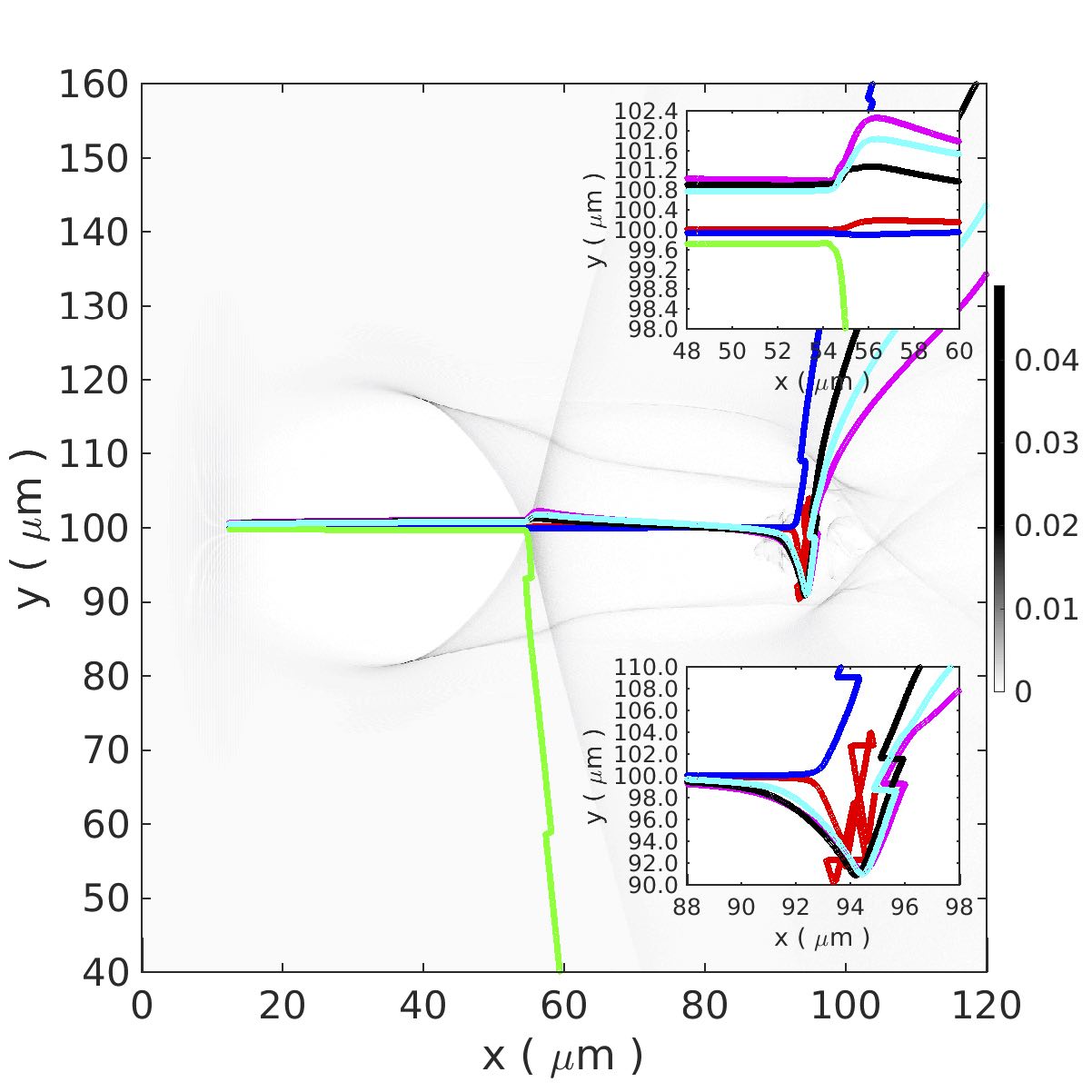}}
\subfigure[]
{\includegraphics[width=7.0cm]{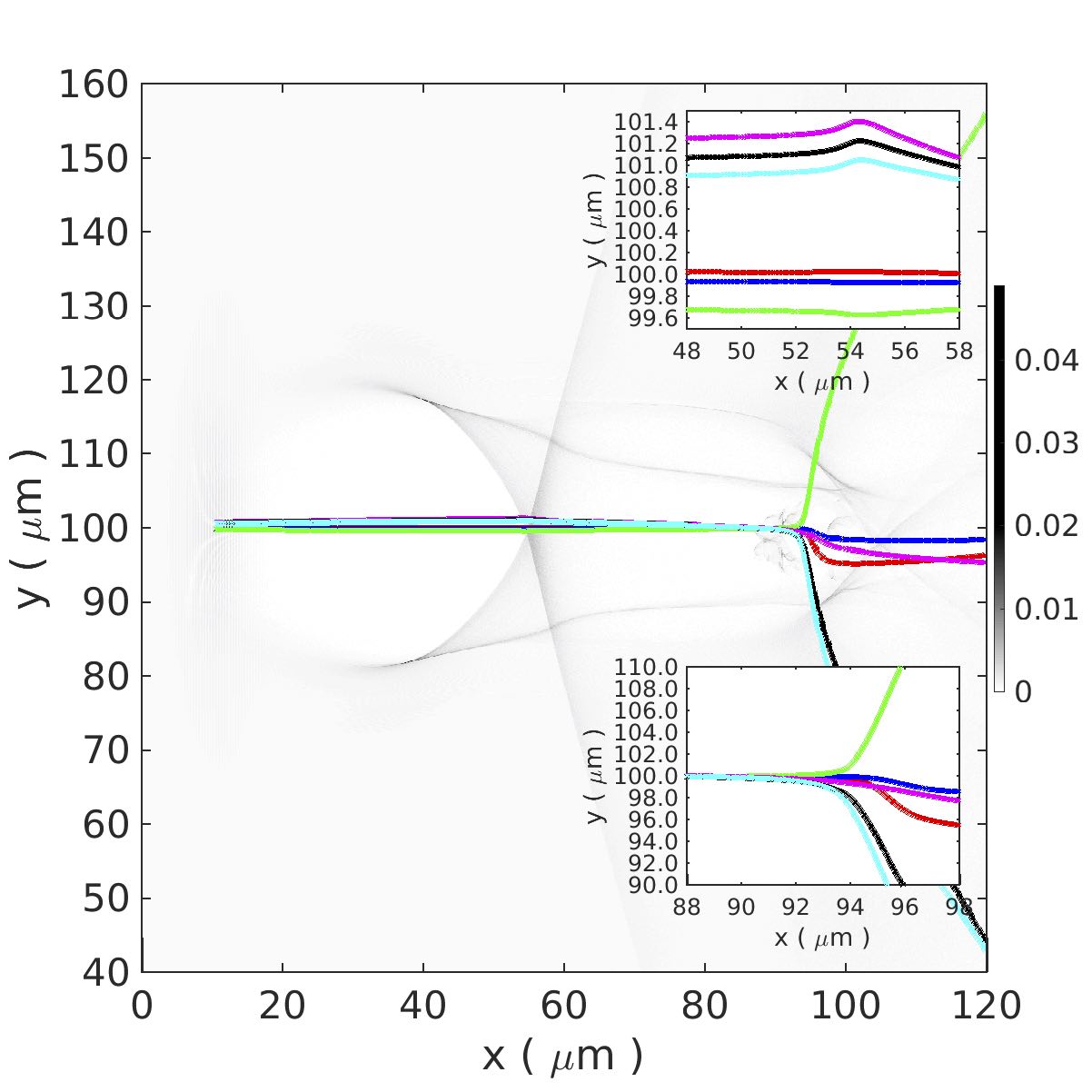}}
\hspace {0.3cm}
\subfigure[]
{\includegraphics[width=7.0cm]{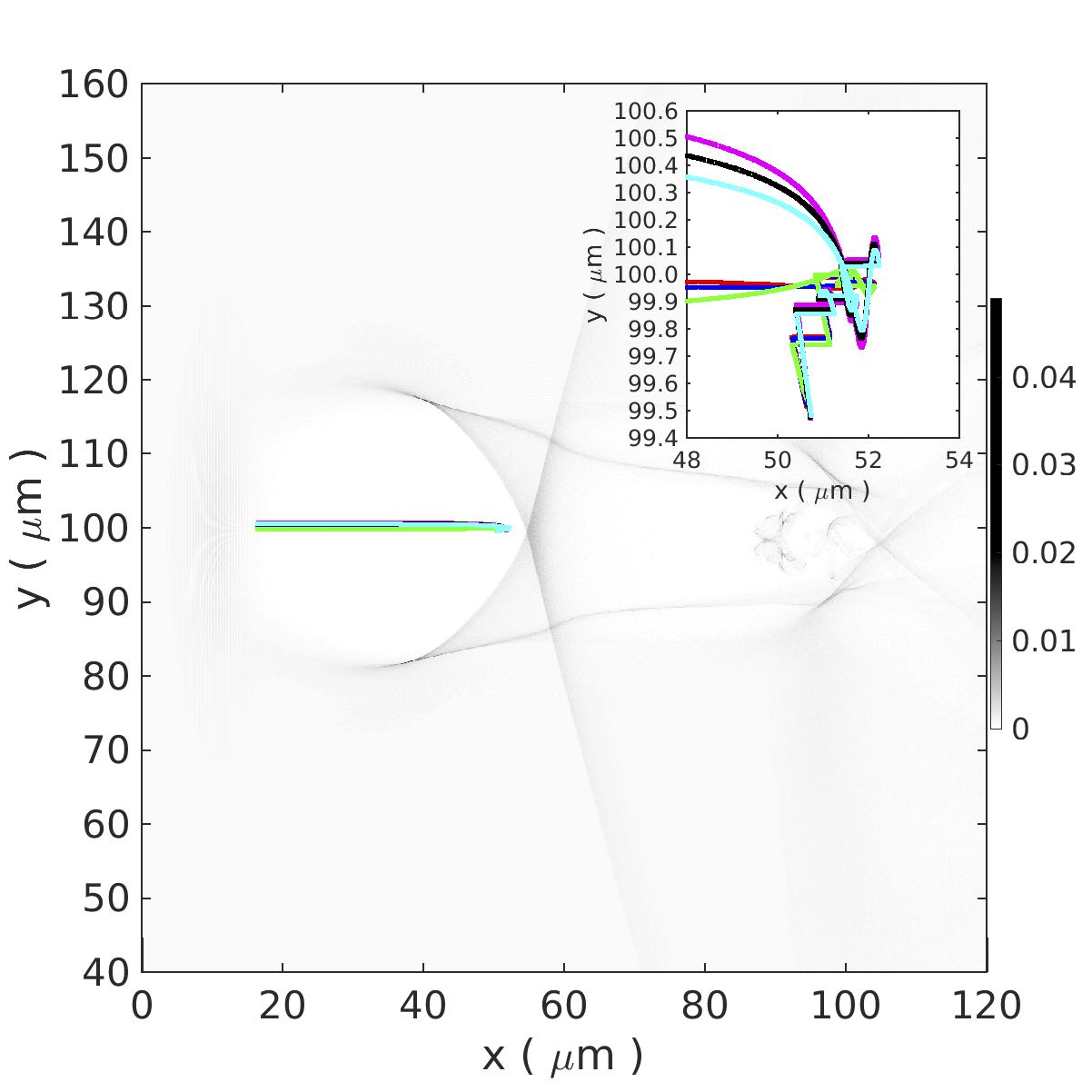}}
\caption{ Trajectories of test probe particles placed at distance $\bigtriangleup x$ from the maximum of the laser pulse with 30 fs duration and intensity $I=3 \times 10^{19} Wcm^{-2}$. (a) $\bigtriangleup x = 0 \ \mu m$, (b) $\bigtriangleup x = -2 \ \mu m$, (c) $\bigtriangleup x = -4 \ \mu m$, and (d) $\bigtriangleup x = 2 \ \mu m$. Different colors corresponds to different transverse positions of the probe particles: green-$\bigtriangleup y = -0.2 \ \mu m$, blue-$\bigtriangleup y = -0.15 \ \mu m$, red-$\bigtriangleup y = -0.1 \ \mu m$, cyan-$\bigtriangleup y = 0.5 \ \mu m$, black-$\bigtriangleup y = 0.6 \ \mu m$,magenta-$\bigtriangleup y = 0.7 \ \mu m$. Inset shows magnified view of different regions in the wave buckets. Laser pulse is propagating from right to left hand side. The colorbar shows electron density normalized by critical density.}
\label{fig.3}
\end{figure}

\begin{figure}[!tb]
\centering
\subfigure[]
{\includegraphics[width=7.0cm]{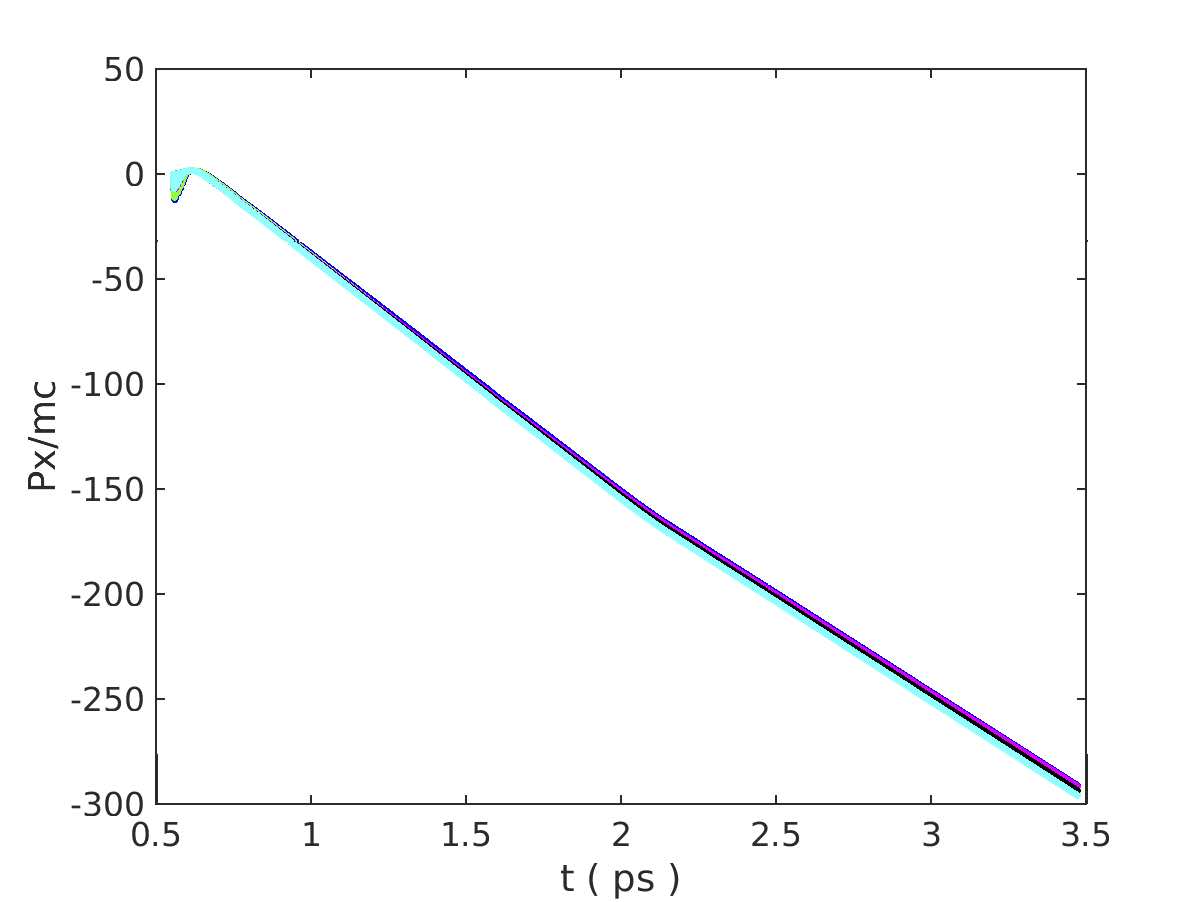}}
\subfigure[]
{\includegraphics[width=7.0cm]{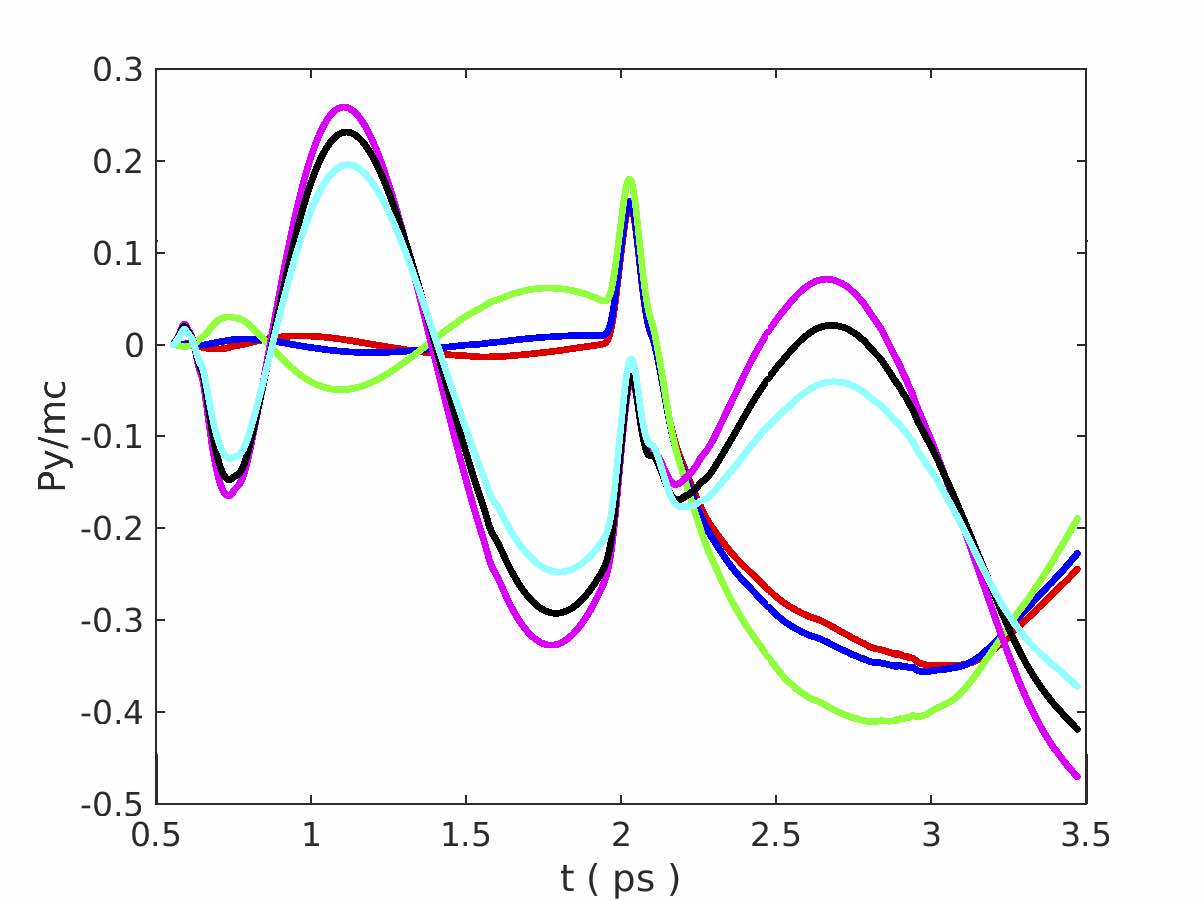}}
\subfigure[]
{\includegraphics[width=7.0cm]{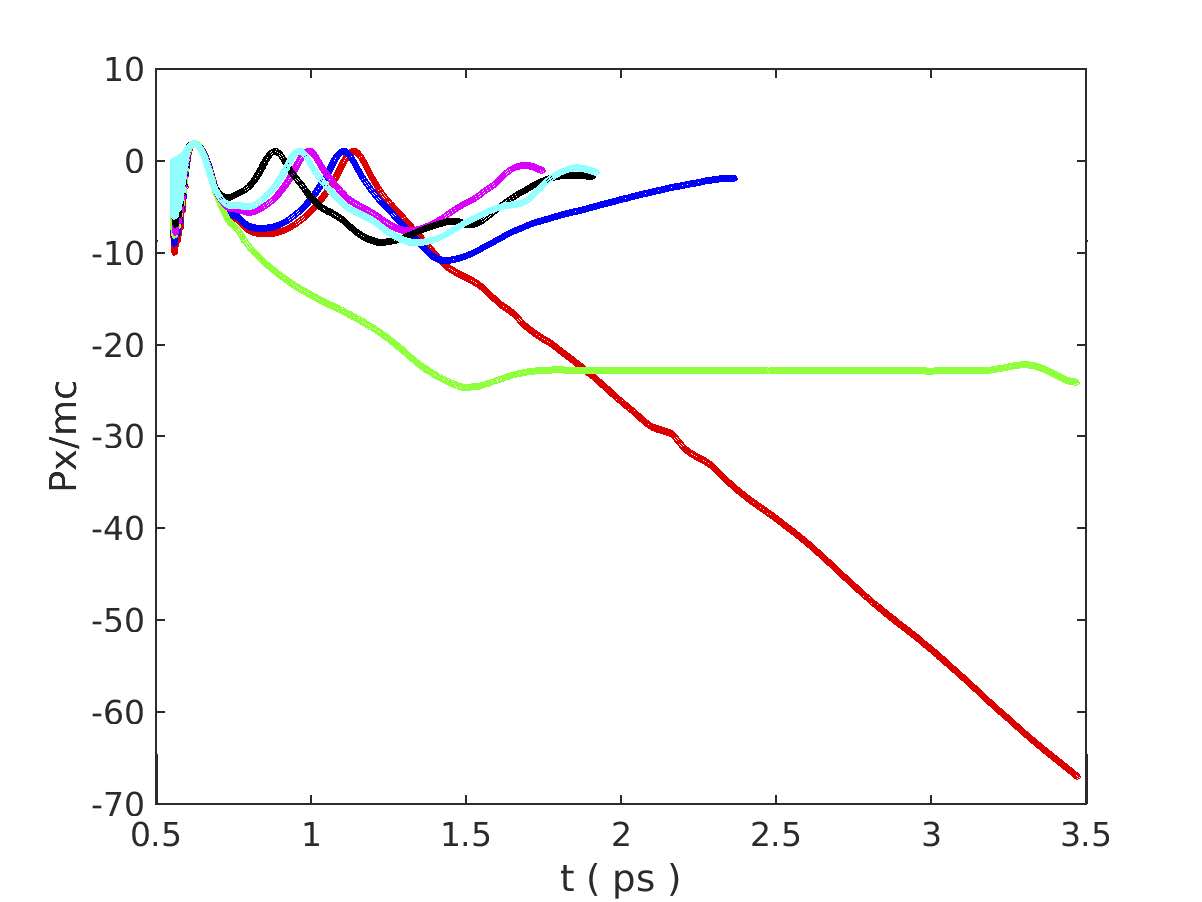}}
\subfigure[]
{\includegraphics[width=7.0cm]{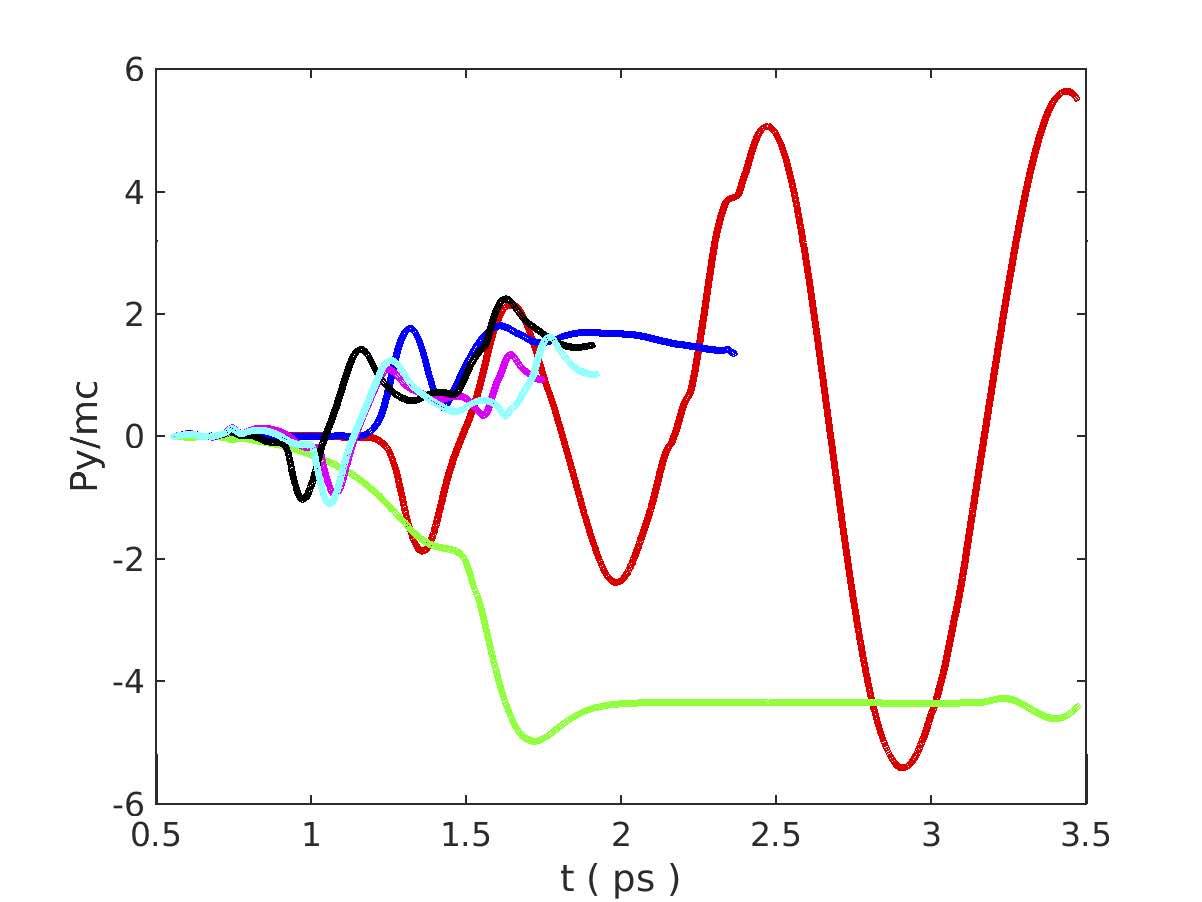}}
\caption{ Evolution of the longitudinal and transverse momentum of the probing particles placed at a distance $\bigtriangleup x$ from the laser pulse with 30 fs duration and intensity $I=3 \times 10^{19} Wcm^{-2}$. (a,b) $\bigtriangleup x = 0 \ \mu m$, and (c,d) $\bigtriangleup x = -2 \ \mu m$. Different colors corresponds to different transverse positions of the probe particles: green-$\bigtriangleup y = -0.2 \ \mu m$, blue-$\bigtriangleup y = -0.15 \ \mu m$, red-$\bigtriangleup y = -0.1 \ \mu m$, cyan-$\bigtriangleup y = 0.5 \ \mu m$, black-$\bigtriangleup y = 0.6 \ \mu m$,magenta-$\bigtriangleup y = 0.7 \ \mu m$.}
\label{fig.4}
\end{figure}

\begin{figure}[!tb]
\centering
\subfigure[]
{\includegraphics[width=12cm]{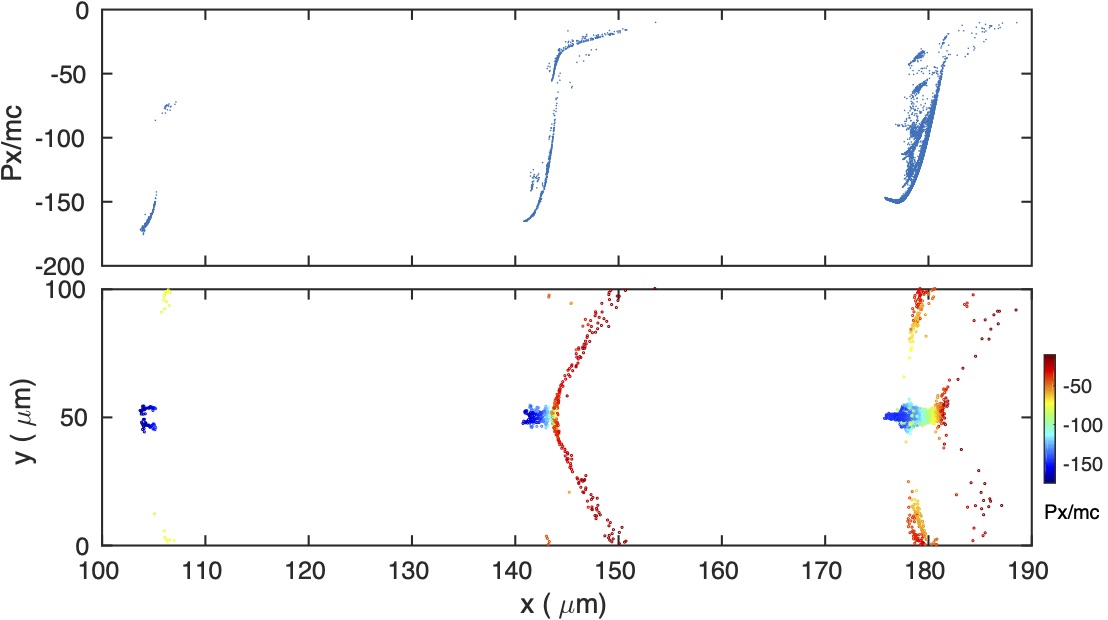}}
\subfigure[]
{\includegraphics[width=12cm]{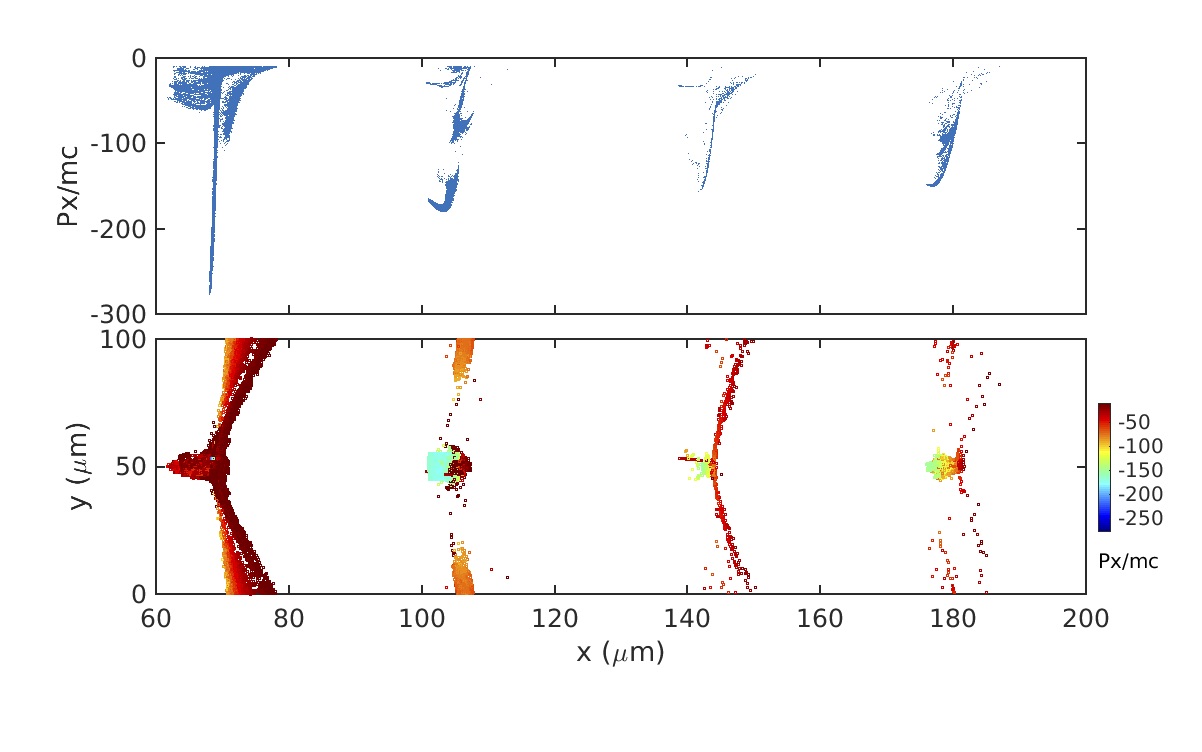}}
\caption{ Spatial distribution of electrons momenta at $t=4.1 \ ps$ in a plasma with $N_{e}=10^{18}cm^{-3}$, and $\bigtriangleup N = 10^{16}cm^{-3}$ irradiated by a laser pulse with $I=3 \times 10^{19} Wcm^{-2}$, $\tau=30 \ fs$. (a) 'plasma' electrons, and (b) 'ionization' electrons. Laser pulse is propagating from right to left hand side.}
\label{fig.5}
\end{figure}

\begin{figure}[!tb]
\centering
\subfigure[]
{\includegraphics[width=8cm]{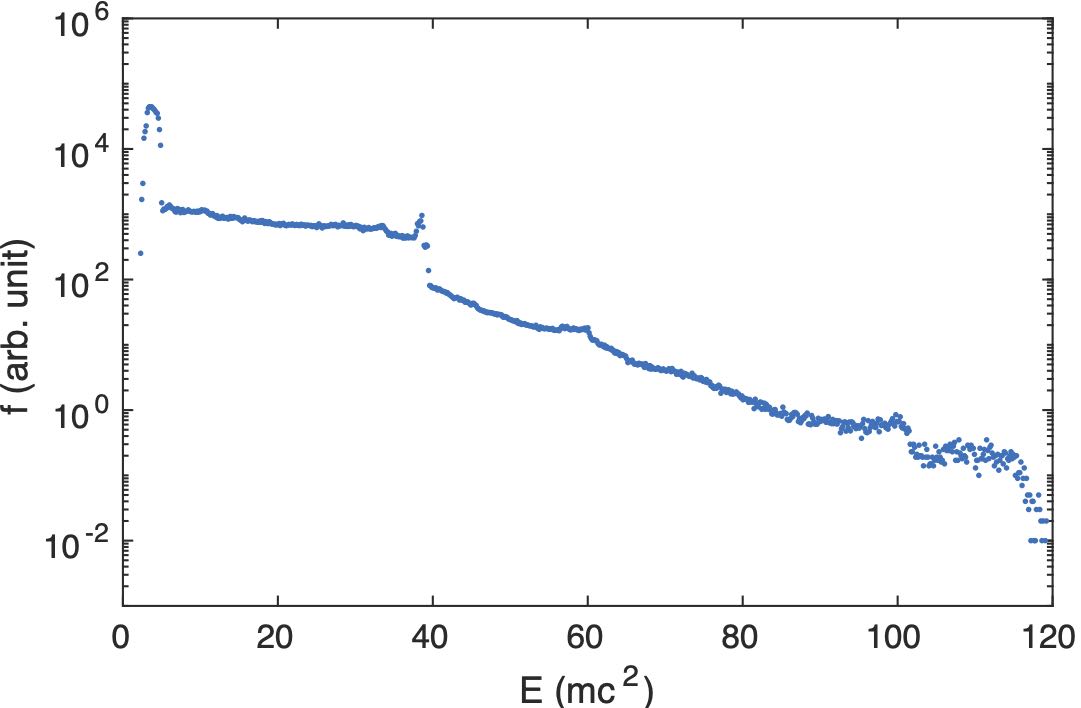}}
\subfigure[]
{\includegraphics[width=8cm]{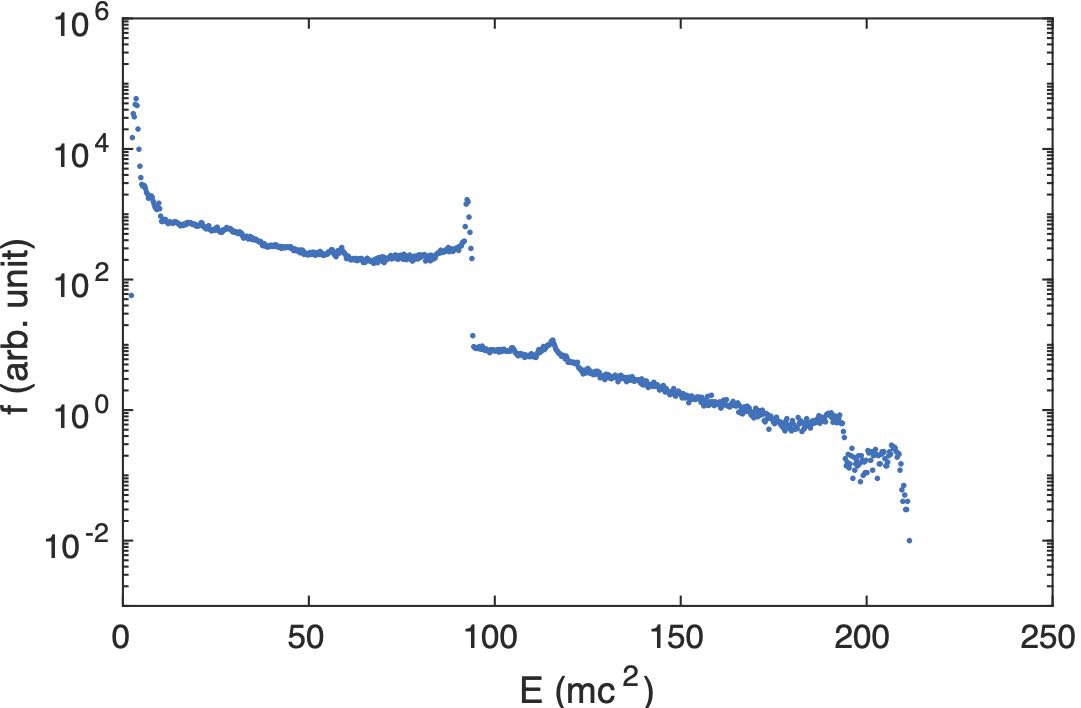}}
\subfigure[]
{\includegraphics[width=8cm]{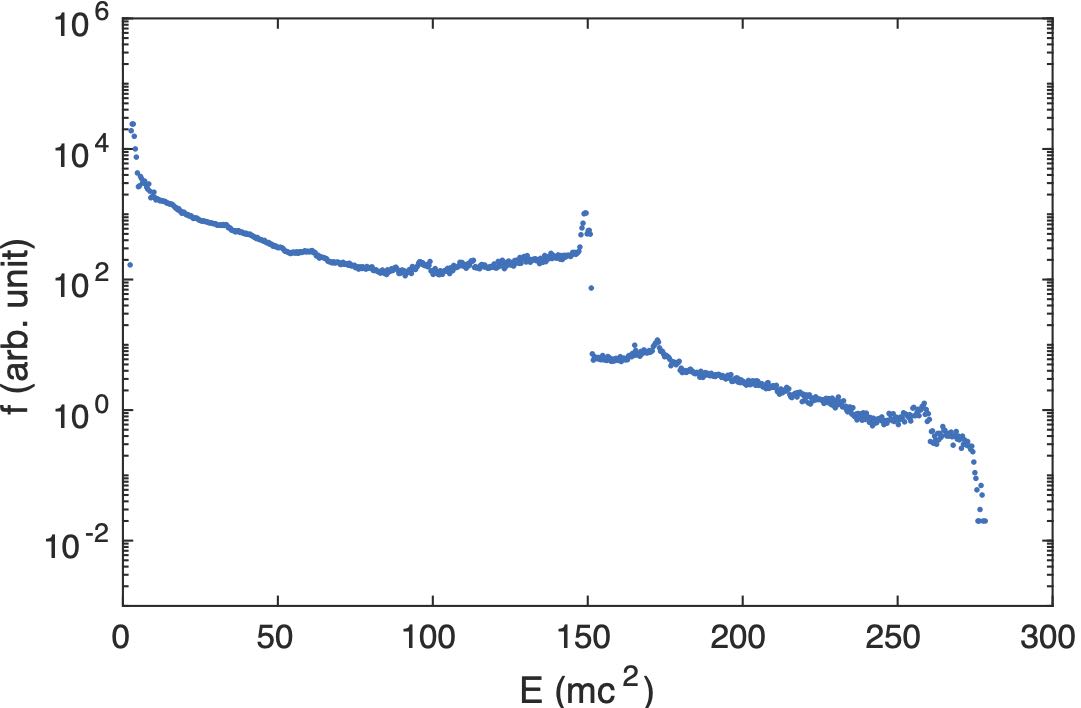}}
\caption{ Evolution of the total energy distribution for a plasma with $N_{e}=10^{18}cm^{-3}$,$\bigtriangleup N=10^{16}cm^{-3}$ irradiated by a laser pulse with $I=3 \times 10^{19} Wcm^{-2}$, $\tau=30 \ fs$ at (a) $t=1.7 \ ps$, (b) $t=2.9 \ ps$, and (c) $t=4.1 \ ps$.}
\label{fig.6}
\end{figure}

\begin{figure}[!tb]
\centering
\subfigure[]
{\includegraphics[width=7cm]{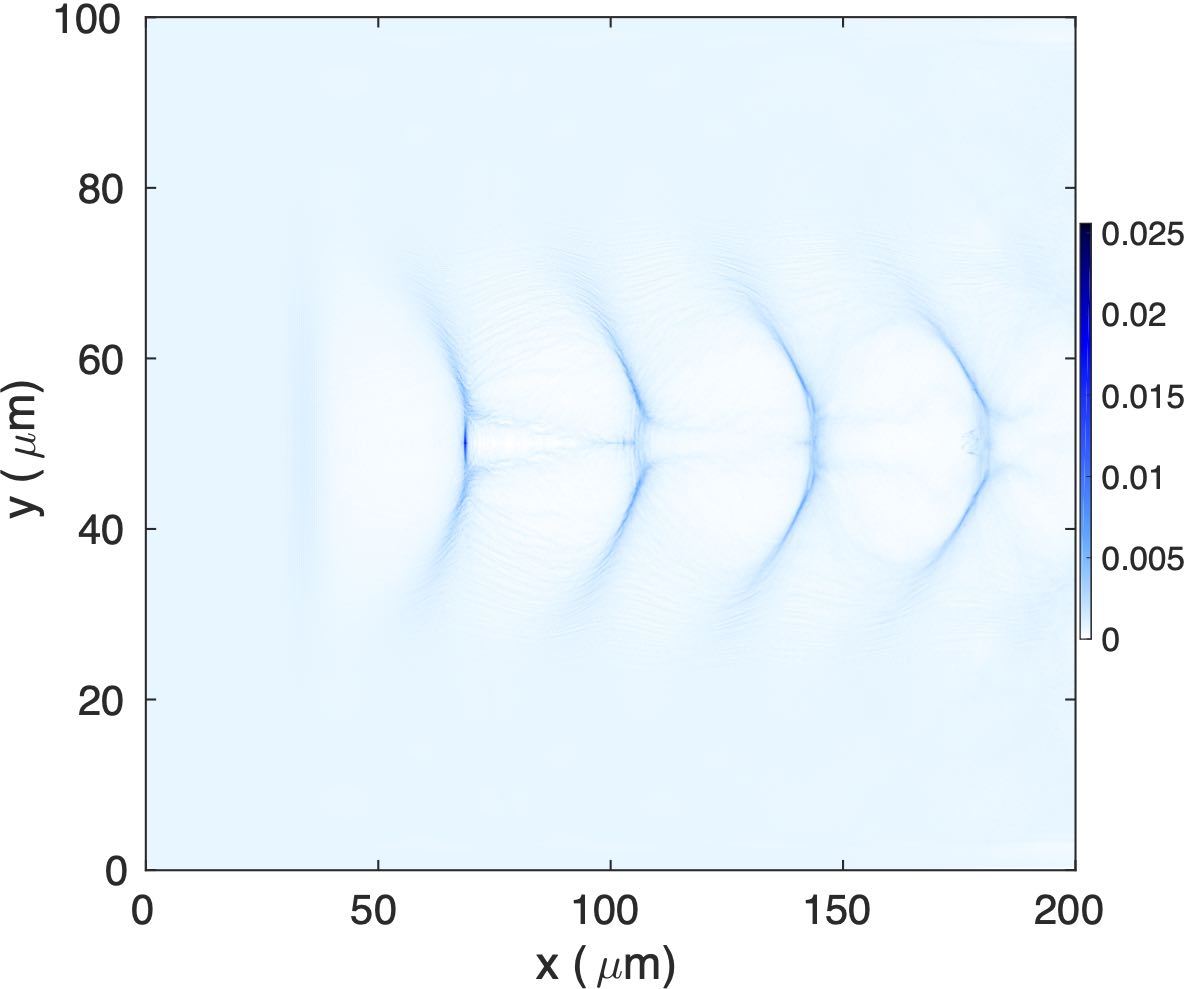}}
\hspace {0.5cm}
\subfigure[]
{\includegraphics[width=6.8cm]{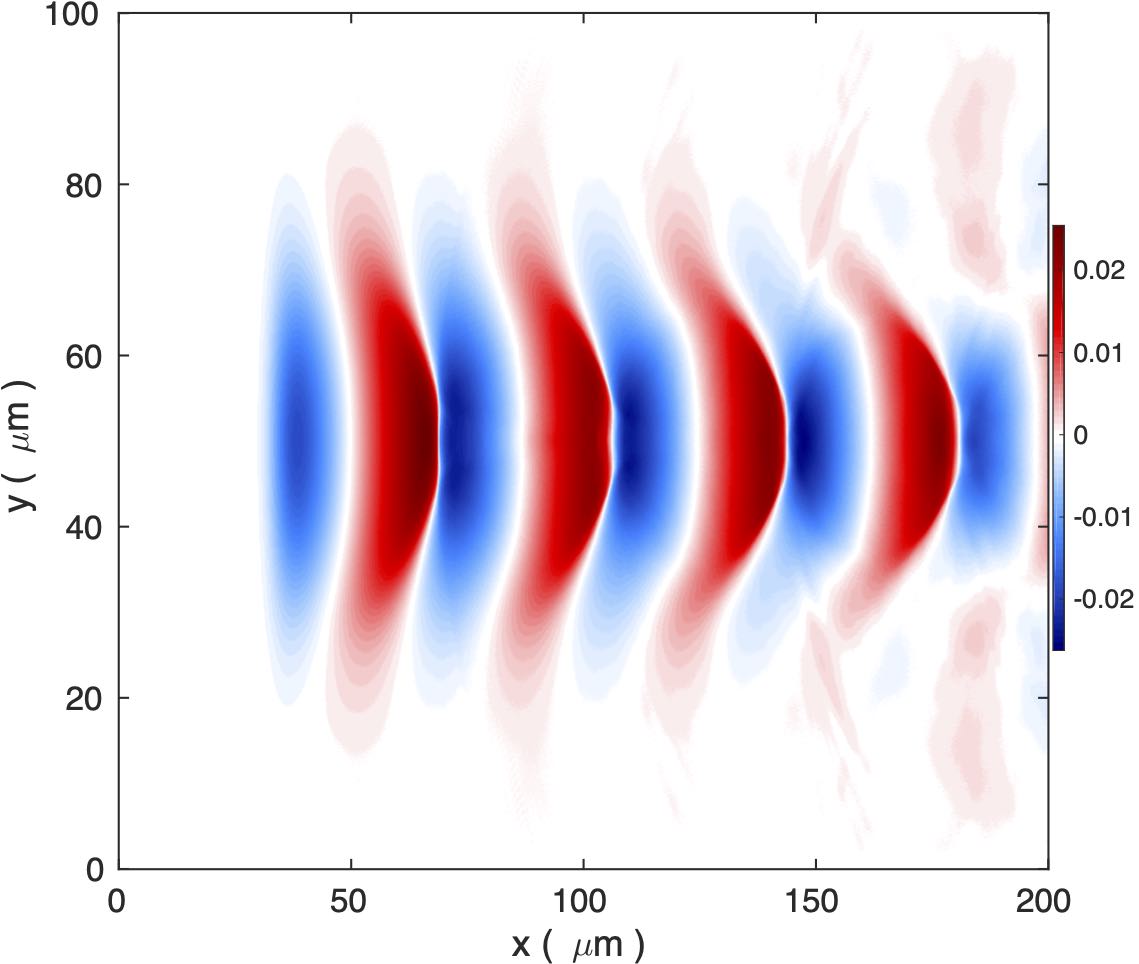}}
\caption{Spatial distribution of (a) electrons density and (b) x-component of the electric field for a plasma with $N_{e}=10^{18}cm^{-3}$, and $\bigtriangleup N = 10^{16}cm^{-3}$ irradiated by a laser pulse with $I=3 \times 10^{19} Wcm^{-2}$, $\tau=30 \ fs$ at $t=4.1 \ ps$. Laser pulse is propagating from right to left hand side. In (a) the colorbar shows electron density normalized by critical density and in (b) the colorbar shows axial field in normalized units ($eEx/m\omega c$).}
\label{fig.7}
\end{figure}

\begin{figure}[!thbp]
\centering
\subfigure[]
{\includegraphics[width=12cm]{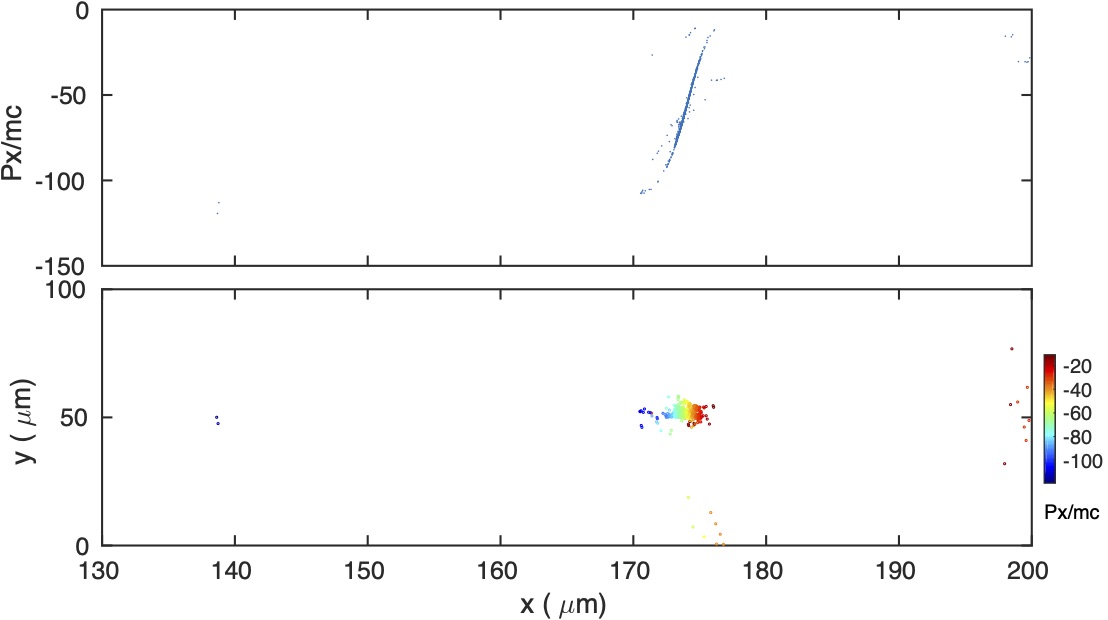}}
\subfigure[]
{\includegraphics[width=13cm]{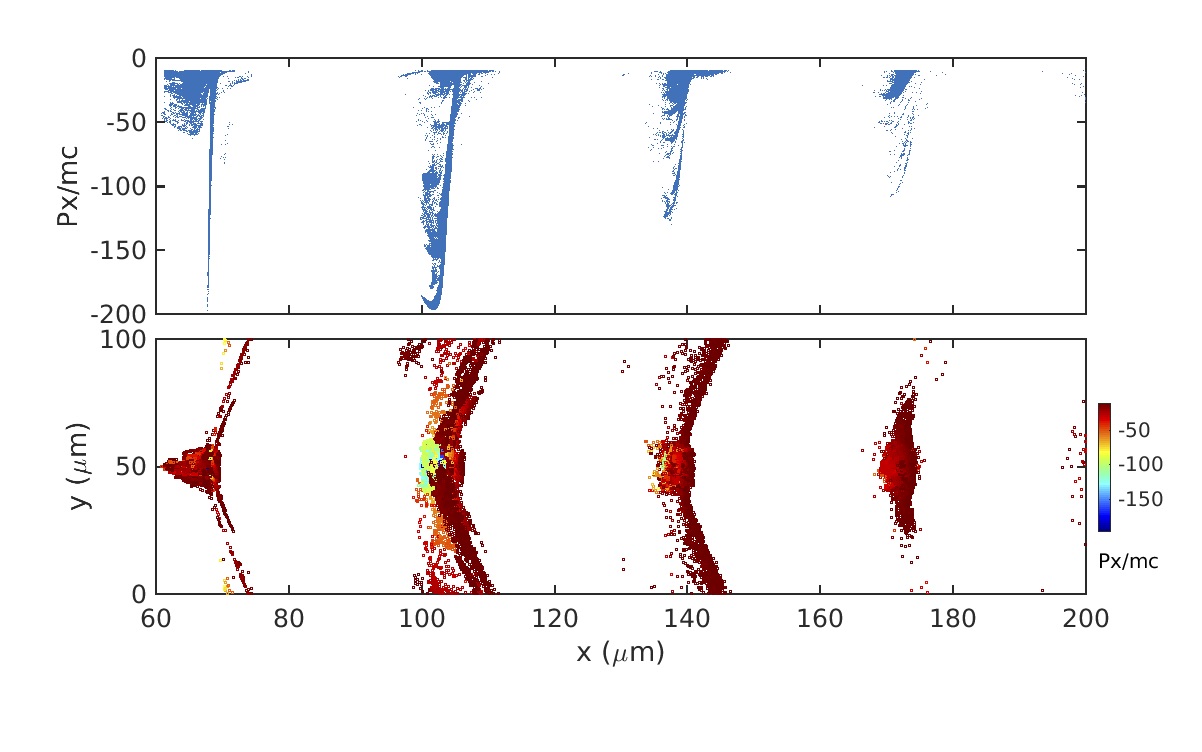}}
\caption{ Spatial distribution of electrons momenta at $t=4.1 \ ps$ in a plasma with $N_{e}=10^{18}cm^{-3}$, and $\bigtriangleup N = 10^{17}cm^{-3}$ irradiated by a laser pulse with $I=3 \times 10^{19} Wcm^{-2}$, $\tau=30 \ fs$. (a) 'plasma' electrons, and (b) 'ionization' electrons. Laser pulse is propagating from right to left hand side.}
\label{fig.8}
\end{figure}

\begin{figure}[!tb]
\centering
{\includegraphics[width=8cm]{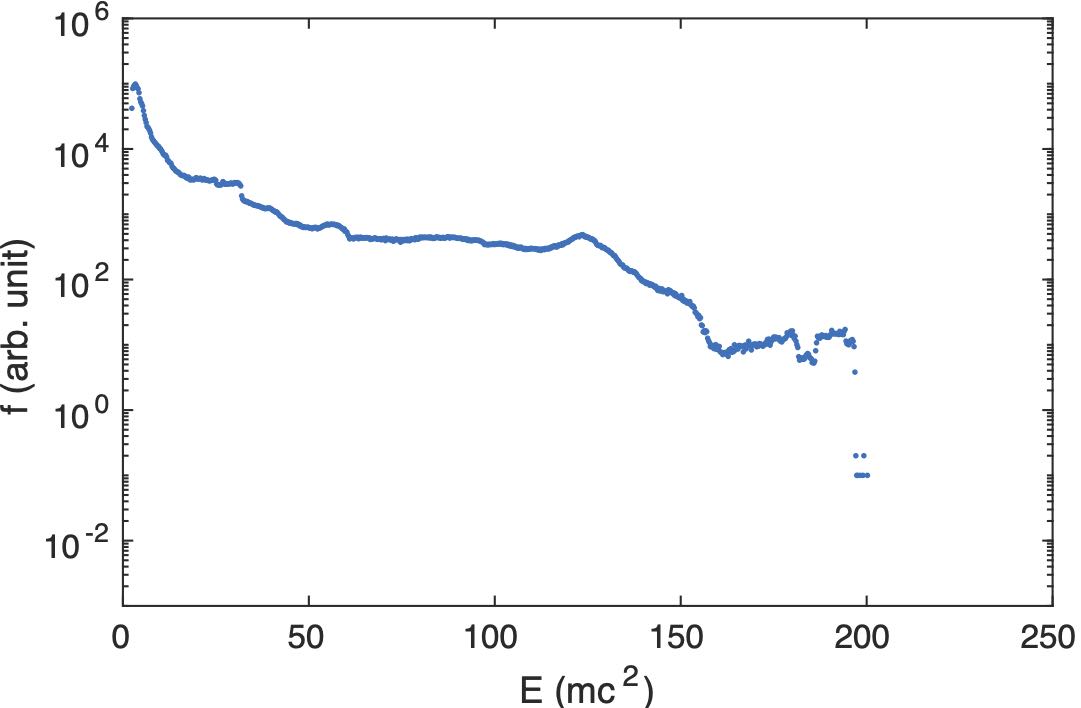}}
\caption{ Total energy distribution for a plasma with $N_{e}=10^{18}cm^{-3}$,$\bigtriangleup N=10^{17}cm^{-3}$ irradiated by a laser pulse with $I=3 \times 10^{19} Wcm^{-2}$, $\tau=30 \ fs$ at $t=4.1 \ ps$.}
\label{fig.9}
\end{figure}

\begin{figure}[!thbp]
\centering
\subfigure[]
{\includegraphics[width=12cm]{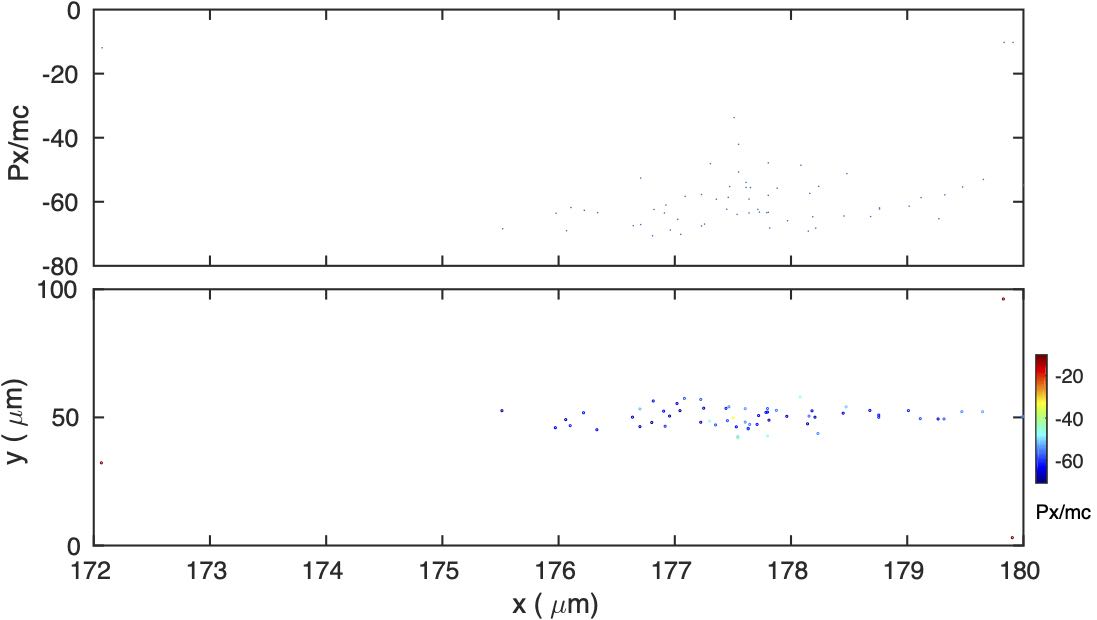}}
\subfigure[]
{\includegraphics[width=12cm]{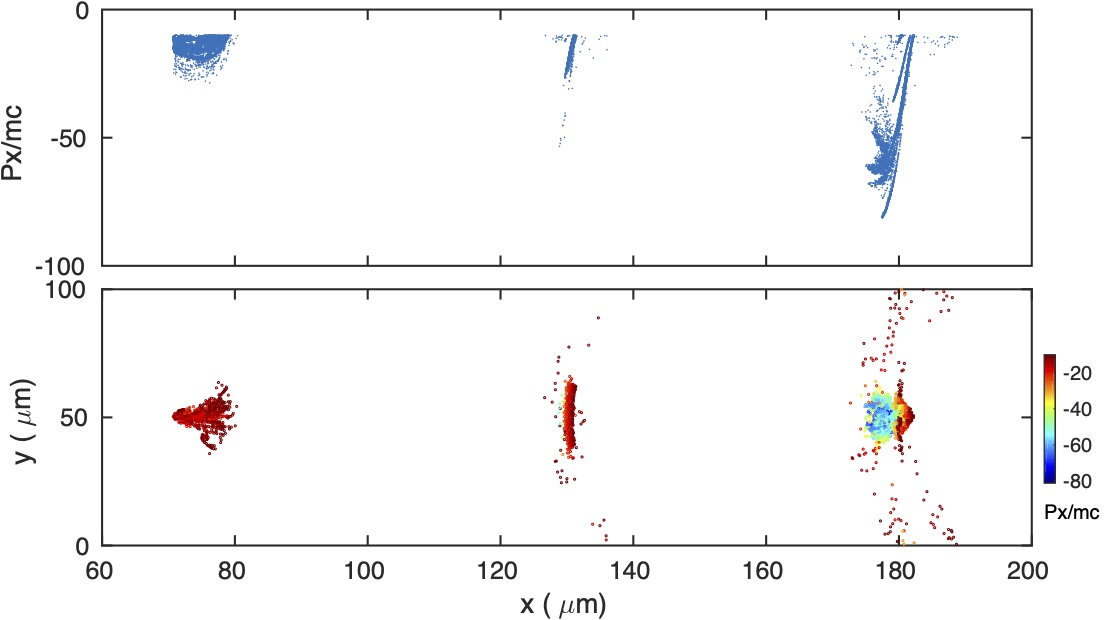}}
\caption{ Spatial distribution of electrons momenta at $t=4.1 \ ps$ in a plasma with $N_{e}=5\times 10^{17}cm^{-3}$, and $\bigtriangleup N = 5\times 10^{16}cm^{-3}$ irradiated by a laser pulse with $I=3 \times 10^{19} Wcm^{-2}$, $\tau=30 \ fs$. (a) 'plasma' electrons, and (b) 'ionization' electrons. Laser pulse is propagating from right to left hand side.}
\label{fig.10}
\end{figure}

\begin{figure}[!tb]
\centering
\subfigure[]
{\includegraphics[width=6.9cm]{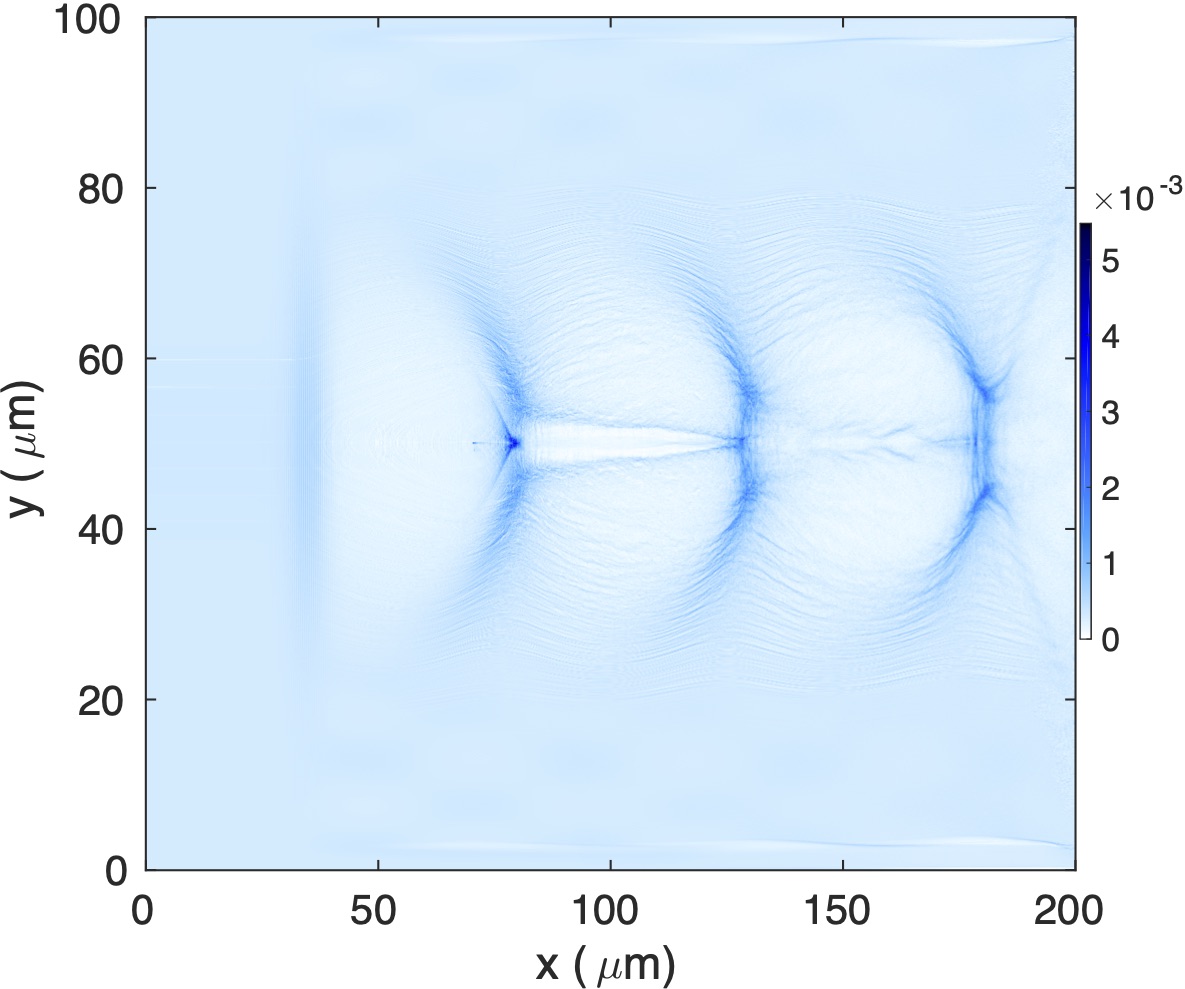}}
\hspace {0.5cm}
\subfigure[]
{\includegraphics[width=6.9cm]{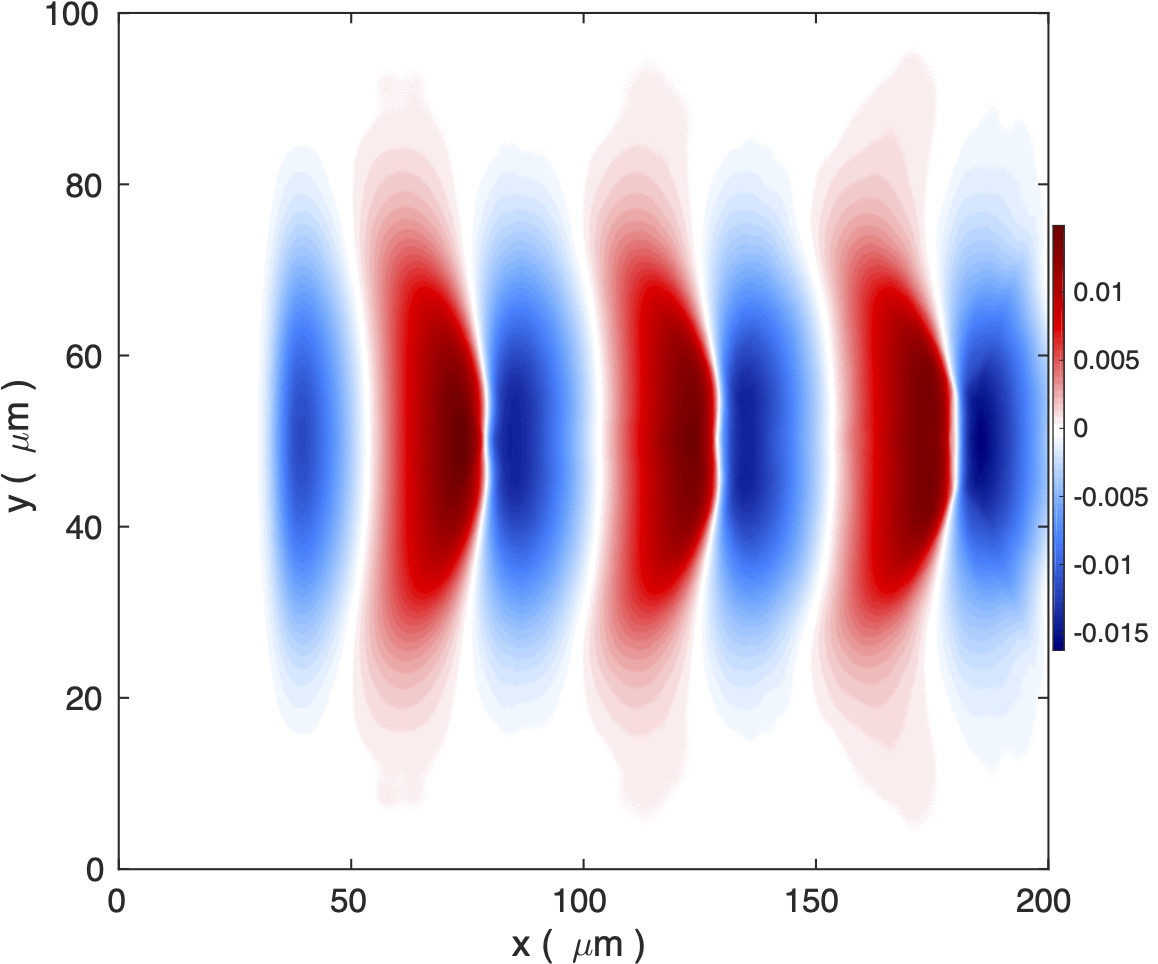}}
\caption{Spatial distribution of (a) electrons density and (b) x-component of the electric field for a plasma with $N_{e}=5\times 10^{17}cm^{-3}$, and $\bigtriangleup N = 5\times 10^{16}cm^{-3}$ irradiated by a laser pulse with $I=3 \times 10^{19} Wcm^{-2}$, $\tau=30 \ fs$ at $t=4.1 \ ps$. Laser pulse is propagating from right to left hand side. In (a) the colorbar shows electron density normalized by critical density and in (b) the colorbar shows axial field in normalized units ($eEx/m\omega c$).}
\label{fig.11}
\end{figure}

\begin{figure}[!tb]
\centering
\subfigure[]
{\includegraphics[width=8cm]{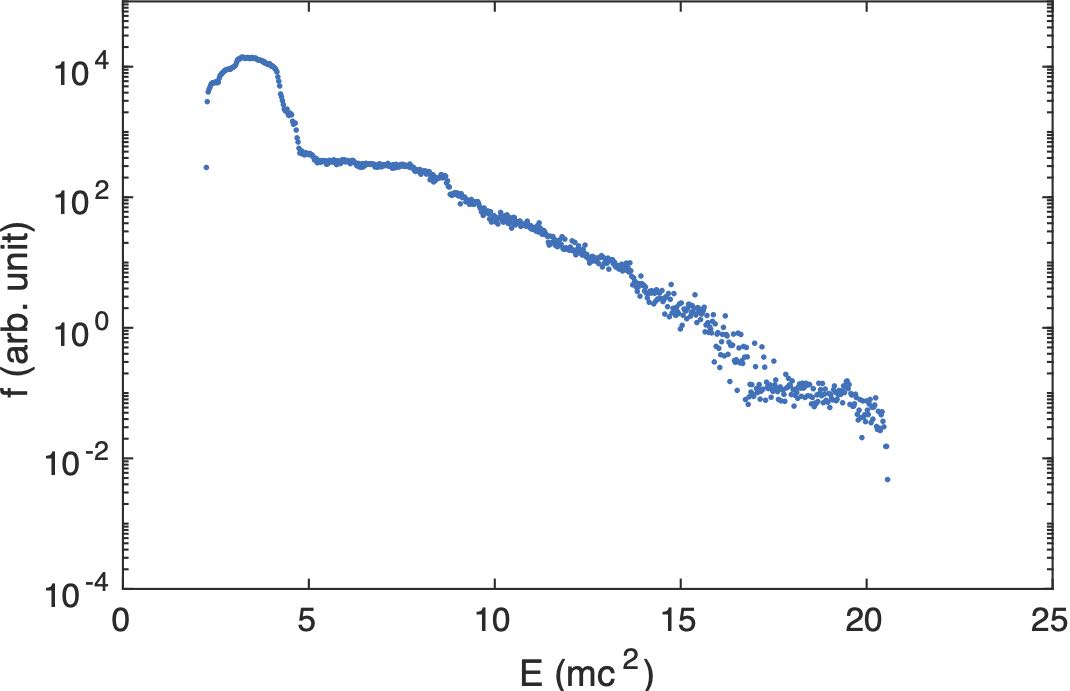}}
\vspace{0.1cm}
\subfigure[]
{\includegraphics[width=8cm]{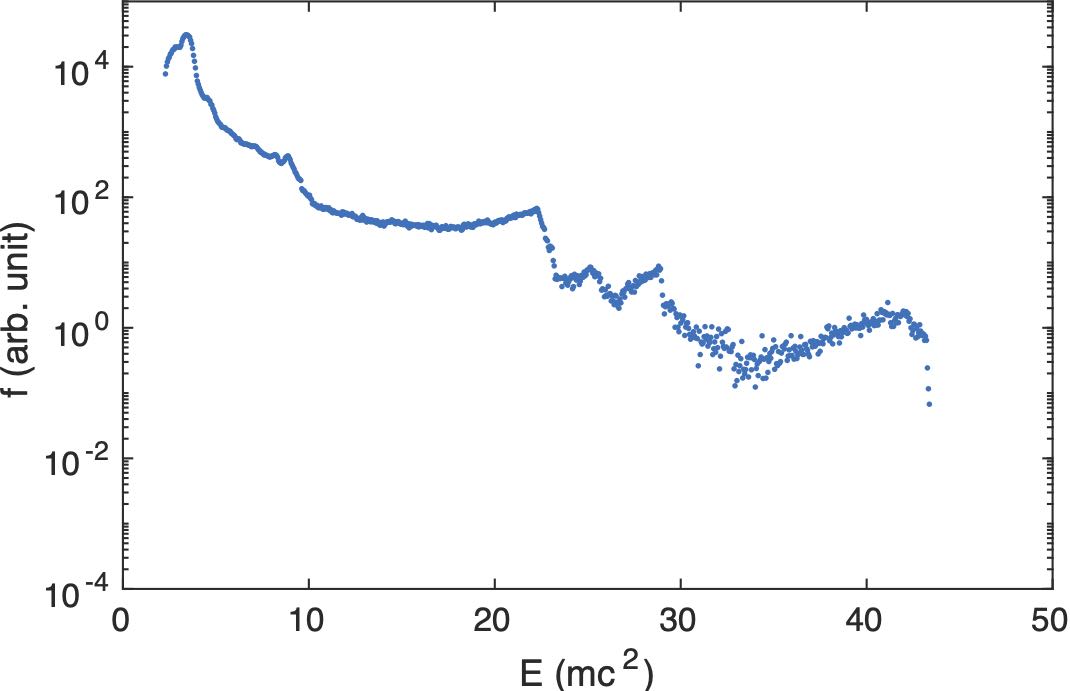}}
\vspace{0.1cm}
\subfigure[]
{\includegraphics[width=8cm]{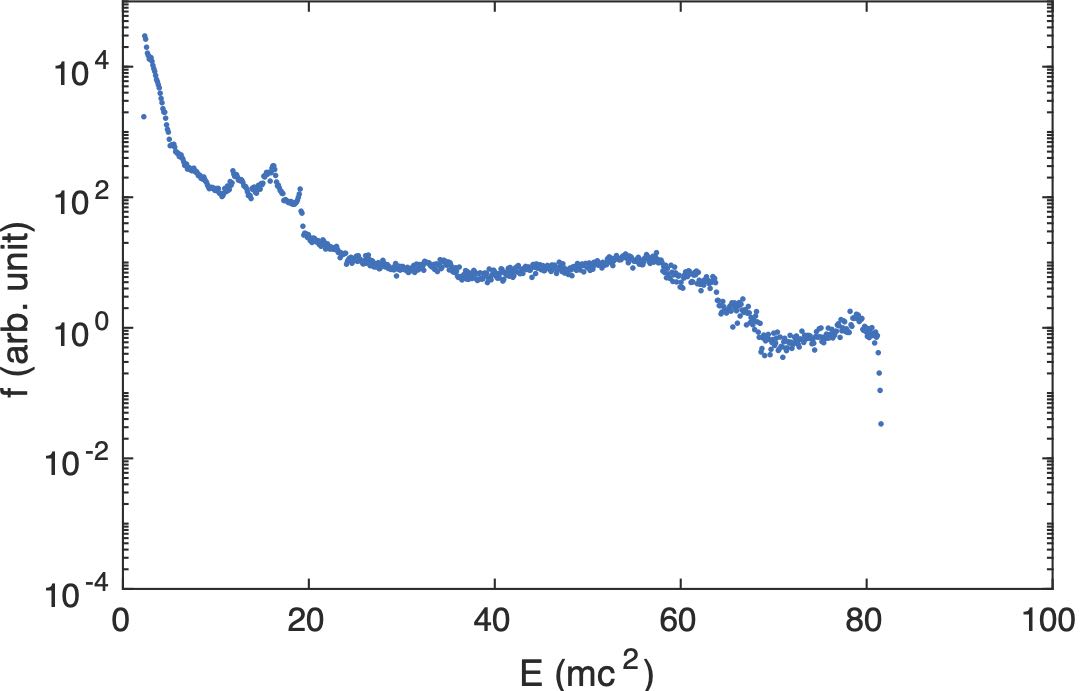}}
\caption{ Evolution of the total energy distribution for a plasma with $N_{e}=5\times 10^{17}cm^{-3}$,$\bigtriangleup N=5\times 10^{16}cm^{-3}$ irradiated by a laser pulse with $I=3 \times 10^{19} Wcm^{-2}$, $\tau=30 \ fs$ at (a) $t=1.7 \ ps$, (b) $t=2.9 \ ps$, and (c) $t=4.1 \ ps$.}
\label{fig.12}
\end{figure}


\end{document}